\documentclass[a4paper,10pt,oneside]{article}

\usepackage{graphicx}%
\usepackage{multirow}%
\usepackage{amsmath,amssymb,amsfonts}%
\usepackage{amsthm}%
\usepackage{mathrsfs}%
\usepackage[title]{appendix}%
\usepackage{xcolor}%
\usepackage{textcomp}%
\usepackage{manyfoot}%
\usepackage{booktabs}%
\usepackage{algorithm}%
\usepackage{algorithmicx}%
\usepackage{algpseudocode}%

\usepackage[utf8]{inputenc}
\usepackage[T1]{fontenc}
\usepackage[english]{babel}

\usepackage{listings}%

\usepackage{enumitem}

\usepackage{bbold}
\usepackage{braket} 

\newtheorem*{postulate}{Postulate}

\title{The classical boundaries of the EPR argument and quantum ontology}
\author{Vincenzo Chilla}
\date{}

\begin{document}

\maketitle

\begin{abstract}

Von Neumann's Hilbert-space formalism of quantum mechanics constitutes a logico-physical theory of observed or measured reality. Imposing the logical constraint of Booleanity, essential for objectively shareable descriptions among observers, reveals the physical meaning of classicality inherently embedded within the formalism itself. Starting from this consideration, the present work reformulates the quantum-classical transition via Hilbert-space classical mechanics (HCM), grounding classicality not in the dynamical limit ($\hbar \to 0$), but in the logical constraint of Booleanity (i.e., the mutual commutativity of preparable states). Within this state-centric framework, applying the Einstein-Podolsky-Rosen (EPR) criterion alongside locality and measurement independence reduces standard quantum mechanics to the HCM model. Thus, the EPR argument reveals not quantum incompleteness, but the implicit classical boundaries of its own premises. To resolve this impasse, we articulate a nuanced quantum ontology grounded in a fundamental structural bipartition between the observational environment and the observed object, which accommodates three categorical distinctions: ontic, processional, and tropos-existential. Building on this, we propose a criterion of objective reality wherein descriptive objectivity is treated as merely a sufficient condition for physical reality. This addresses the historical Bohr-Einstein ambiguity, enabling the quantum formalism to ontologically unify objective measured phenomena and non-objective observed interference within a context-dependent framework.



\end{abstract}

\section{Introduction}
\label{intro}

Influenced by Spinoza's monism, pre-quantum physics tended to view nature as the ``Absolute One''~\cite{GreneNails1986,Gabbey1996,Jammer1999}, wherein the ``attribute of extension'' grounds the so-called ``elements of physical reality'' detectable through observation. Within this deterministic framework, the observer is an integral part of reality who must remain passive, perceiving natural perfection without influencing the rigid course of events~\cite{BrewerWatkins2012,Melamed2017}. Consequently, a non-invasive observation allows for the formulation of an ontologically \emph{complete} theory, whereas perceived perturbation implies inadequacy. This form of strong scientific physical realism establishes the epistemological basis of the Einstein-Podolsky-Rosen (\emph{EPR}) \emph{argument} against the completeness of quantum mechanics~\cite{EPR1935}, marking the culmination of the Einstein-Bohr debate~\cite{Bohr1949,JahnertLehner2022,Paty2022}. 

The famous argument unfolds through the following logical steps.
\begin{enumerate}
\item A necessary ``condition of completeness'' for a physical theory is established as follows (\emph{EPR condition}): ``\emph{Every element of the physical reality must have a counterpart in the physical theory.''}
\item Since the description of physical reality provided by the formalism of the wave function does not allow for simultaneous knowledge of incompatible observables, two irreconcilable physical interpretations -- formulated as a pair of contrapositive material implications -- arise from the EPR condition in step 1 (\emph{EPR dilemma}):
\begin{itemize}
\item if incompatible observables correspond to simultaneous elements of reality, then quantum mechanics is ontologically incomplete -- i.e., quantum probability reflects epistemic uncertainty;
\item if quantum mechanics is ontologically complete, then incompatible observables do not correspond to simultaneous elements of reality -- i.e., quantum probability reflects ontic indeterminacy.
\end{itemize}
\item A ``criterion of physical reality'', intended to apply independently of the theoretical framework, is stated as follows (\emph{EPR criterion}): ``\emph{If, without in any way disturbing a system, we can predict with certainty (i.e., with probability equal to unity) the value of a physical quantity, then there exists an element of physical reality corresponding to this physical quantity.''}
\item Under the assumptions of local realism and measurement independence,
a thought experiment involves two previously interacting particles that are now spatially separated (\emph{EPR experiment}). By performing independent measurements of either position or momentum on the first particle\footnote{One can choose any pair of incompatible observables, e.g., two orthogonal components of the spin~\cite{Bohm1951,BohmAharonov1957}.} and applying the EPR criterion in step 3 to the second undisturbed particle, it is inferred that both incompatible observables correspond to simultaneous elements of physical reality for the second system.
\item Based on the EPR experiment in step 4, the EPR dilemma in step 2 is finally resolved, leading to the conclusion that quantum mechanics is ontologically incomplete (\emph{EPR conclusion}); thus, quantum probability encodes epistemic uncertainty.
\end{enumerate}
Despite the EPR conclusion, the authors leave open the question of whether a complete formulation of quantum theory exists, although they consider such a formulation to be possible. This position justifies the program of completing quantum mechanics that Einstein pursued throughout his life~\cite{Einstein1936,Pais1982}.

In his immediate response to the EPR argument~\cite{Bohr1935a}, Bohr did not challenge steps 1 and 2 presented above; rather, he targeted the EPR criterion in step 3, remarking that ``[it] contains an essential ambiguity when it is applied to problems of quantum mechanics.'' He clarified this ``ambiguity'' by invoking the principle of \emph{complementarity}~\cite{Bohr1935b}, which denies the possibility of jointly defining conjugate quantities independently of the experimental arrangement. On this basis, Bohr rejected the claim that quantum mechanics is ontologically incomplete and maintained that quantum probability must be interpreted as reflecting ontic indeterminacy (see step 2). Viewed through this lens, the ambiguity of the EPR criterion becomes clear: it implicitly assumes an epistemic reading of the wave function~\cite{HarriganSpekkens2010}, whereas a properly quantum criterion of physical reality, grounded in complementarity, requires an ontic interpretation~\cite{PuseyBarrettRudolph2012}. Consequently, reality conditions cannot be formulated in context-independent classical terms~\cite{Gleason1957,KochenSpecker1967}.

This raises the question: to what type of physical reality does the wave function correspond, and how is it linked to the elements of physical reality invoked by the EPR condition? Given that these elements are explicitly represented in the theory, they necessarily constitute \emph{objective} physical reality -- that is, phenomena that can be described by an observer in a \emph{logical} and \emph{communicable} (i.e., unambiguous) manner~\cite{Barad2022}. Consequently, the wave function formalism must be capable of accounting for them. Bohr, however, regarded it as unduly restrictive to assume that the wave function must encompass all and \emph{only} such objective elements; rather, it must also formally represent aspects of physical reality that cannot be expressed unambiguously -- e.g., the interference pattern in the double-slit experiment, which, in contrast to an objective single spot located on the screen, defies a definite description within a single measurement context~\cite{Folse1985}. These aspects constitute what may be termed \emph{non-objective} physical reality.\footnote{Importantly, here ``non-objective'' does not imply ``subjective''. This point clarifies the misconception of ``subjectivism'' often superficially attributed to Bohr's thought.} In other words, contrary to the EPR presupposition, the wave function formalism provides a complete description of physical reality precisely because \emph{it also accommodates non-objective aspects within the experimental framework}. The epistemological implications are profound: \emph{whereas EPR treats descriptive objectivity as equivalent to physical reality, Bohr regarded it as merely a sufficient condition for the latter -- i.e., a criterion -- not its exhaustive definition}.

Notwithstanding his meticulous -- and at times subtle -- critique of the EPR criterion, Bohr did not offer in his rejoinder an explicit, concise statement that could serve as a direct analogue and alternative to the classical notion of physical reality. For this reason, in the present work, we aim to fill this gap by examining how quantum ontology diverges from the classical one; to this end, we introduce a control model: \emph{Hilbert-space classical mechanics} (\emph{HCM}). HCM is directly derived from the standard quantum postulates, as formulated in von Neumann's axiomatic framework~\cite{vNeumann1955,Luders1950}, by appending a suitable \emph{postulate of classicality} that requires all \emph{physically preparable states} of the system under observation to commute. We stress that the postulate of classicality concerns the commutativity of physical states, not of observables. This conceptual distinction is crucial in our framework. Although the literature already contains well-established formulations of classical mechanics within Hilbert space~\cite{Koopman1931,KoopmanvNeumann1932,Bondar2012}, HCM has the advantage of preserving the standard mathematical structure of quantum mechanics while imposing a commutativity constraint on the state space, thereby transforming it into a classical theory. Moreover, while standard treatments of the classical limit rely on $\hbar \to 0$ to recover a deterministic phase space, the present state-centric approach ensures a Boolean logical structure independent of dynamical parameters. This framework enables a conceptual unification that facilitates a systematic comparison between classical and quantum ontologies. HCM is also well-suited for the unified description of the measurement apparatus and the measured object within the experimental process, in accordance with the prescription of classicality required for the former in the Copenhagen interpretation of quantum mechanics~\cite{Howard1994}.

The joint application of HCM and the EPR argument enables a shift away from the traditional focus on the nonlocality of physical reality~\cite{Bell1964}, allowing for the explicit identification of the category error incurred when classical predicates are applied to quantum entities. The resulting quantum ontology thereby emerges as more nuanced than its classical counterpart, acknowledging further ontological distinctions within physical reality that standard classical ontology overlooks:\footnote{The terms \emph{processional} and \emph{tropos-existential} are deliberately adapted from established metaphysical vocabulary~\cite{Kenny2006,FernandezMoujan2024}, serving a heuristic purpose: to articulate ontological distinctions that standard physical formalism tends to flatten. \emph{Processional} evokes the dynamic constitution of being, resonating with the Thomistic distinctions between substance and existence; \emph{tropos-existential} captures the modal structure of actualization, echoing the Aristotelian framework of ``potentia'' and ``actus''~\cite{Strumia2021}.}
\begin{enumerate}
\item \emph{processional}, formalized through the independence between state and observable that standard classical mechanics lacks;
\item \emph{tropos-existential}, formalized through quantum probability rather than classical probability. 
\end{enumerate}

The paper is organized as follows. In Section~\ref{hcm}, we introduce the postulate of classicality and justify it logically by analyzing classical theory within the Hilbert-space framework that arises when classical logic is embedded in the standard quantum postulates. We then provide a description of the formal notions of state, observable, probability, and time, highlighting their correspondence with the standard phase-space formulation of classical mechanics. In Section~\ref{hcmonto}, we examine the ontological status of HCM, revisiting the EPR argument from this perspective. We show that, once the quantum postulates are accepted -- even if only as phenomenological rather than fundamental statements -- the EPR criterion and the associated experiment necessarily lead to HCM. That is, the program of completing quantum mechanics, as originally advocated in the EPR paper, could only yield classical mechanics. In Section~\ref{quonto}, we compare classical and quantum ontologies as they arise from the unified Hilbert-space formalism underlying both. We then formulate a criterion of \emph{objective} reality that is structurally and logically analogous to the EPR criterion, but treats descriptive objectivity as merely a sufficient (rather than necessary and sufficient) condition for physical reality. This criterion captures the core of quantum ontology. In Section~\ref{concl}, we conclude by summarizing our findings and presenting some final considerations and remarks. Finally, Appendix~\ref{gloss} contains a glossary that collects the main ontological terms used in the present work; this vocabulary is not an external imposition, but rather emerges \emph{a posteriori} directly from the Hilbert-space formalism constrained by the postulate of classicality.

\section{The formalism of HCM}
\label{hcm}

We start by presenting the postulates for a foundational formulation of \emph{non-relativistic} quantum mechanics in a standard and traditional Hilbert-space formalism~\cite{vNeumann1955,Luders1950}. For simplicity, we assume a finite-dimensional Hilbert space, restrict measurements to ideal, projective, repeatable (von~Neumann-L\"uders) processes, and work strictly within this regime. These assumptions allow us to isolate the essential logical structure of the theory, deferring technical generalizations (continuous spectra, unbounded operators, POVMs, or relativistic extensions) to more advanced treatments~\cite{PeresTerno2004,Strocchi2008}.

\begin{enumerate}[label=\Roman*.]

\item \textbf{Physical states.} The state of a physical system is completely described by a \emph{statistical operator} (or \emph{density operator}) $\rho$. This operator acts on a complex, separable Hilbert space $\mathcal{H}$ associated with the system. The statistical operator must be \emph{Hermitian}, \emph{positive semi-definite}, and of \emph{unit trace} ($\operatorname{Tr} (\rho) = 1$).\footnote{This formalism encompasses both \emph{pure} states, for which $\rho = \ket{\psi}\bra{\psi}$ (where $\ket{\psi} \in \mathcal{H}$ is a normalized vector), and \emph{mixed} states.}

\item \textbf{Observables.} Each observable of the physical system corresponds to a \emph{Hermitian operator} $\hat{O}$ acting on the system's Hilbert space $\mathcal{H}$. The possible outcomes of a measurement of the observable are given by the eigenvalues of the corresponding operator.\footnote{While the formalism associates Hermitian operators with real-valued measurement outcomes, physical constraints or superselection rules may restrict which operators correspond to experimentally accessible observables.}

\item \textbf{Measurement.} 
\begin{enumerate}
\item A measurement of an observable $\hat{O}$ on a system in state $\rho$ yields one of the eigenvalues $o_k$ of $\hat{O}$. The probability of obtaining the result $o_k$ is given by the \emph{Born rule}:
    \[
    p(o_k) = \operatorname{Tr}(\rho \, \mathbb{P}_k),
    \]
    where $\mathbb{P}_k$ represents the \emph{orthogonal projector} onto the eigenspace of $\hat{O}$ corresponding to $o_k$.  
    
\item If the measurement yields the result $o_k$, the state of the system changes instantaneously according to the \emph{L\"uders projection rule}:
    \[
    \rho' = \frac{\mathbb{P}_k \rho \, \mathbb{P}_k}{\operatorname{Tr}(\rho \, \mathbb{P}_k)}.
    \]
\end{enumerate}

\item \textbf{Time evolution.} The state of a \emph{closed} (isolated) system evolves in time according to a \emph{unitary transformation}. In the density operator formalism, this evolution is governed by:
\[
\rho(t) = U(t) \, \rho(0) \, U^\dagger(t),
\]
where $U(t)$ is the unitary time-evolution operator. For a time-independent Hamiltonian $\hat{H}$, $U(t) = \exp(-i\hat{H}t/\hbar)$.\footnote{This formulation corresponds to the \emph{Schr\"odinger picture}. For time-dependent Hamiltonians, $U(t)$ requires a time-ordered exponential; for open systems, unitary evolution is replaced by a non-unitary dynamical map, typically governed by a Markovian master equation.}

\item \textbf{Composite systems.} The Hilbert space $\mathcal{H}_{AB}$ of a \emph{composite physical system}, consisting of two subsystems $A$ and $B$, is the \emph{tensor product} of the individual Hilbert spaces: $\mathcal{H}_{AB} = \mathcal{H}_A \otimes \mathcal{H}_B$. If the subsystems are prepared \emph{independently}, the joint state is the tensor product of the individual states: $\rho_{AB} = \rho_A \otimes \rho_B$.\footnote{The tensor product structure naturally accommodates both product states and \emph{entangled} (non-separable) states, providing the mathematical framework for quantum correlations.}

\item \textbf{Identical parts.} The state of a system composed of indistinguishable parts must be either \emph{totally symmetric} under the exchange of any two parts (\emph{bosons}) or \emph{totally antisymmetric} under such exchange (\emph{fermions}).\footnote{This symmetrization postulate applies to all indistinguishable quantum subsystems, regardless of whether they are elementary or composite. It underlies \emph{quantum statistics} (Bose-Einstein and Fermi-Dirac) and the \emph{Pauli exclusion principle}.}

\end{enumerate}
To bridge quantum and classical mechanics within the Hilbert-space framework, we introduce the required classicality condition in the following section.

\subsection{Postulate of classicality}
\label{class}

Before proceeding with an explicit formulation of the \emph{postulate of classicality}, it is necessary to clarify what is meant by ``classical theory''. The answer to this question is inherently \emph{epistemological}: \emph{a physical theory is defined as classical if the experimental propositions constituting it obey Boolean logic}. Consequently, the term ``classical'' does not refer directly to ontological commitments, but rather to the \emph{logical} framework through which an analytic observer \emph{knows} the state of a physical system. This perspective aligns with the approach originally adopted by Birkhoff and von Neumann in presenting the foundations of \emph{quantum logic}~\cite{BirkhoffvNeumann1936}, where the physical knowledge of the observer is represented by experimental propositions of the form 
\begin{equation}
\mathfrak{p}_\Delta = \{ \text{the system is in the \emph{state} } \ket{x}\bra{x} \text{ with \emph{measured} value } x \in \Delta \},
\label{exprop}
\end{equation}
and $x$ represents a list of common eigenvalues for a \emph{complete set of commuting observables} (\emph{CSCO}) taking values in $\Delta$. The experimental logic is then mapped onto the quantum formalism, as defined by the postulates stated above, via the correspondence
\begin{equation}
\mathfrak{p}_\Delta \longleftrightarrow \mathbb{E}_\Delta \doteq \sum_{x \in \Delta} \mathbb{P}_x,
\label{qlogic}
\end{equation}
where $\mathbb{P}_x$ denotes the orthogonal projector $\ket{x}\bra{x}$, and $\mathbb{E}_\Delta$ can be termed a \emph{propositional projector}. 

Within this framework, the truth value of the proposition $\mathfrak{p}_\Delta$ for a system prepared in state $\rho$, denoted $v_\rho(\mathfrak{p}_\Delta)$, is \emph{formally} defined as the probability of obtaining a ``yes'' outcome in the corresponding ``yes-no'' measurement~\cite{Jauch1968,FoulisBennett1994}: 
\begin{equation}
v_\rho(\mathfrak{p}_\Delta) \doteq \operatorname{Tr}(\rho \mathbb{E}_\Delta).
\label{truth_v}
\end{equation}
This definition ensures general consistency with the Born rule (Postulate~III) and the spectral decomposition theorem. If $v_\rho(\mathfrak{p}_\Delta) = 1$, $\mathfrak{p}_\Delta$ is true for the state $\rho$; if $v_\rho(\mathfrak{p}_\Delta) = 0$, $\mathfrak{p}_\Delta$ is false; otherwise, the truth value is indeterminate and the corresponding physical property is logically \emph{indefinite}. 
Crucially, while the truth value of $\mathfrak{p}_\Delta$ is determined by the state $\rho$, the proposition itself encodes a physical property (represented by the projector $\mathbb{E}_\Delta$), rather than a generic observable. In this sense, $\mathfrak{p}_\Delta$ constitutes a logical assertion about the \emph{state} that is independent of any specific measurement context, and $\mathbb{E}_\Delta$ should \emph{not} be identified with a physical observable~\cite{BeltramettiCassinelli1981}. Therefore, one must avoid identifying the truth value formally defined in~\eqref{truth_v} with the expectation value associated with the measurement of $\mathbb{E}_\Delta$ according to the Born rule, as if the latter were a physical observable in the sense of Postulate~II.\footnote{Birkhoff and von Neumann~\cite{BirkhoffvNeumann1936} originally advanced what they termed a ``not unnatural conjecture,'' namely that all self-adjoint operators -- including propositional projectors -- correspond to physical observables. This identification, while historically influential, is neither necessary nor universally tenable and has been extensively criticized on both operational and foundational grounds~\cite{Peres1984, Peres2003}.}

The correspondence established in~\eqref{qlogic} endows the set of logical propositions with the structure of an \emph{orthomodular lattice}, in which the distributive law does not generally hold. This algebraic framework is fundamentally distinct from that of Boolean logic, which is characterized as a \emph{complemented and distributive lattice}. However, \emph{a subset of quantum propositions forms a Boolean algebra if and only if the corresponding propositional projectors mutually commute}~\cite{BirkhoffvNeumann1936}. Moreover, \emph{the mutual commutativity of propositional projectors is equivalent to the mutual commutativity of the physically preparable states associated with them}. In fact, if propositional projectors mutually commute, the commutativity of the preparable states directly follows from the spectral decomposition of the state and the linearity of the commutator. Conversely, to prove that the mutual commutativity of preparable states implies the commutativity of propositional projectors, consider two arbitrary propositional projectors $\mathbb{E}_{\Delta_1}$ and $\mathbb{E}_{\Delta_2}$, and the associated \emph{states} $\rho_{1}$ and $\rho_{2}$ defined as
\begin{equation}
\rho_{i} = \frac{\mathbb{E}_{\Delta_i}}{\operatorname{Tr}(\mathbb{E}_{\Delta_i})}.
\label{prep_s}
\end{equation}
(The case $\mathbb{E}_{\Delta_i} = \mathbb{0}$ is trivial; we therefore assume $\mathbb{E}_{\Delta_i} \neq \mathbb{0}$.) Each state in~\eqref{prep_s} is \emph{preparable} and corresponds to a \emph{statistical mixture with uniform probabilities} $1/N_i$, where $N_i=\operatorname{Tr}(\mathbb{E}_{\Delta_i}) > 0$. By hypothesis, the preparable states $\rho_{1}$ and $\rho_{2}$ commute, so $[\rho_{1}, \rho_{2}] = 0$. Since $\rho_{i} = \frac{1}{N_i}\mathbb{E}_{\Delta_i}$, the linearity of the commutator implies $[\mathbb{E}_{\Delta_1}, \mathbb{E}_{\Delta_2}] = N_1 N_2 [\rho_{1}, \rho_{2}] = 0$. Because $\mathbb{E}_{\Delta_1}$ and $\mathbb{E}_{\Delta_2}$ are arbitrary, it follows that all propositional projectors commute.

The explicit equivalence between the mutual commutativity of propositional projectors and that of the physically preparable states associated with them constitutes the conceptual bridge between logic and quantum mechanics.\footnote{This equivalence assumes operational completeness: the set of physically preparable states spans the algebra of experimentally accessible propositions, ensuring that every non-zero projector admits a physical preparation.} Just as quantum logic reduces to classical logic when all experimental propositions are required to form a Boolean algebra, quantum mechanics reduces to classical mechanics when all physically preparable states are required to commute. This finally permits the formulation of the following \emph{postulate of classicality}.
\begin{postulate}[\textbf{Classicality}]
All preparable states of a physical system, described by statistical operators on the system's Hilbert space $\mathcal{H}$, mutually commute. 
\end{postulate}
The inclusion of the preceding postulate among the quantum postulates stated above establishes a \emph{superselection rule} that enables the formalization of classical mechanics within the Hilbert-space framework -- i.e., HCM -- and derives it directly from quantum mechanics. We underscore the privileged role played by physically preparable states: their assumed mutual commutativity constitutes the direct physical expression of classicality, as opposed to the commutativity of observables. In what follows, we examine in greater detail the relationship and conceptual distinction between these two notions within the context of HCM.

\subsection{States and observables}
\label{stob}

The first consequence of classicality, i.e., the mutual commutativity of physically preparable states, is the existence of a common orthonormal eigenbasis $\{ \ket{\gamma_i} \}$ -- as guaranteed by the spectral theorem for commuting operators -- which we shall term the \emph{classical basis} of the Hilbert space of states $\mathcal{H}$ (see Appendix~\ref{gloss}). It follows that \emph{the classical basis constitutes the unique physically admissible orthonormal basis}, up to permutations of the basis vectors and phase factors, for the description of the system's states. Indeed, the orthogonal projector onto any linear superposition of two or more classical basis vectors does not correspond to a physically preparable state. To prove this, consider a normalized superposition state $\ket{\psi} = \sum_k c_k \ket{\gamma_k}$ containing at least two non-vanishing coefficients, e.g., $c_i \neq 0$ and $c_j \neq 0$ with $i \neq j$. The corresponding pure-state density operator is $\rho_\psi = \ket{\psi}\bra{\psi}$; we evaluate its commutator with the \emph{classical state} $\mathbb{P}_i= \ket{\gamma_i} \bra{\gamma_i}$:
\begin{align}
[\rho_\psi, \mathbb{P}_i] &= \ket{\psi}\bra{\psi}\ket{\gamma_i}\bra{\gamma_i} - \ket{\gamma_i}\bra{\gamma_i}\ket{\psi}\bra{\psi} \nonumber \\
&= \sum_{k \neq i} \left( c_i^* c_k \ket{\gamma_k}\bra{\gamma_i} - c_i c_k^* \ket{\gamma_i}\bra{\gamma_k} \right).
\end{align}
Since there exists at least one index $j \neq i$ such that $c_j \neq 0$, the term corresponding to $k=j$ is non-zero, implying $[\rho_\psi, \mathbb{P}_i] \neq 0$. But if $\rho_\psi$ were physically preparable, classicality would require $[\rho_\psi, \mathbb{P}_i] = 0$, as $\mathbb{P}_i$ corresponds to a physically preparable pure state; thus, $\rho_\psi$ cannot be physically preparable. From Postulate~I, \emph{only statistical mixtures of classical states, $\rho= \sum_i p_i \ket{\gamma_i}\bra{\gamma_i}$ with $p_i \geq 0$ and $\sum_i p_i = 1$, are physically preparable in HCM}.

Now, suppose we observe the system and measure one of its observables~$\hat{O}$. If the measured value corresponds to the eigenvalue~$o_k$, according to Postulate~III, the system is in the classical state~$\mathbb{P}_k$, and the \emph{eigenstate-eigenvalue link}~\cite{Fine1973,Albert1992}, expressed by the relation $\hat{O} \mathbb{P}_k = o_k \mathbb{P}_k$, holds. Consequently, $\hat{O}$ \emph{commutes with the classical state and, by extension, with every physically preparable state}~$\rho$. Thus, like these states, \emph{all observables mutually commute and are represented by operators that are diagonal in the classical basis}. As a result, the contextual distinction inherent to the quantum framework is lost. For this reason, the physical state -- being represented by a Hermitian operator -- can be easily conflated with an observable in the sense of Postulate~II, effectively collapsing the physical distinction between them.\footnote{While the previous result on the mutual commutativity of observables could serve as the postulate of classicality, ensuring the completeness of HCM would demand the further, unjustified assumption that the orthogonal projectors associated with the common eigenbasis qualify as observables according to Postulate~II.} In other words, although classicality is defined in terms of the commutativity of physically preparable states, paradoxically, one could be led to regard the observable as the main and general physical notion, thereby reducing the state to a formal and constrained construct.\footnote{This is consistent with ordinary phase-space classical mechanics, where the position $Q$ and the momentum $P$ of a particle constitute paradigmatic examples of observables, and the ordered pair $(Q, P)$ is the state. The notion of state is then a formal aid, whose physical meaning resides solely in the observables $Q$ and $P$ from which it is formally constructed.} Therefore, the failure to properly distinguish between the physical state and the observable can be a symptom of an underlying classical setting. 

According to Postulate~III, the expectation value of an observable $\hat{O}$ for a system prepared in the state $\rho = \sum_i p_i \ket{\gamma_i}\bra{\gamma_i}$ is given by
\begin{equation}
    \langle \hat{O} \rangle_\rho = \operatorname{Tr}(\rho \hat{O}) = \sum_i p_i o_i,
    \label{exp_v}
\end{equation}
where $o_i$ are the eigenvalues of $\hat{O}$. Any change of basis in the Hilbert space $\mathcal{H}$, leading to a basis $\{ \ket{\phi_i} \}$ whose vectors correspond to states that cannot be physically prepared, leaves the expectation value in~\eqref{exp_v} invariant. Within the HCM framework, such a basis transformation does not correspond to a change in the experimental context in the quantum-mechanical sense; consequently, there is no operational means to distinguish the associated states from classical ones. In other words, non-preparable states are epistemologically inaccessible and empirically indistinguishable from classical states; the corresponding ``coherence terms'' possess only formal significance, lacking any observable content that could be revealed via interference effects.

Based on the foregoing, a Hermitian operator on $\mathcal{H}$ that is diagonal in the non-classical basis $\{ \ket{\phi_i} \}$, given by $A = \sum_i a_i \ket{\phi_i}\bra{\phi_i}$, \emph{does not} qualify as an observable in the sense of Postulate~II.\footnote{Note the absence of a circumflex on the symbol $A$, highlighting its purely mathematical nature as an operator rather than a physical observable.} Indeed, if it did, Postulate~III would entail that obtaining a measurement outcome $a_k$ would update the system's state to the classical state $\ket{\gamma_k}\bra{\gamma_k}$. However, an immediately repeated measurement of the same observable $A$ would not yield a deterministic result, but only a probabilistic outcome with conditional probabilities $|\braket{\gamma_k | \phi_i}|^2$. This would therefore constitute a ``noisy'' and non-repeatable measurement that fails to reveal the system's pre-existing state with certainty. Thus, $A$ cannot correspond to a genuine physical observable of the system. This leads to the following \emph{condition of observability}, which, while inherent to the HCM framework due to the uniqueness of the context-defining classical basis, also extends to the quantum case: \emph{within a fixed observational context, a Hermitian operator acting on the system's Hilbert space qualifies as a physical observable only if it is diagonal in the context-defining (classical) basis -- i.e., only if it commutes with every physically preparable state}. In other words, altering the reference basis so as to render an operator non-diagonal is not merely a formal mathematical exercise, but corresponds to a \emph{shift} toward a complementary context (cf. Appendix~\ref{gloss}).\footnote{This stringent condition drastically reduces the algebra of admissible observables, which is precisely the intended feature of HCM as a control model: it isolates the logical skeleton of a theory in which all measurable quantities share a single, context-independent reference frame.} As a corollary, within a \emph{fixed} observational context, the post-measurement state update prescribed by the L\"uders rule does not describe a physical transformation of the system, but rather constitutes a purely epistemic revision aimed solely at updating the observer's state of knowledge. A case in point is the classical measuring apparatus in the Copenhagen interpretation, where the L\"uders rule constitutes nothing more than an epistemic update upon pointer observation.\footnote{Throughout this work, \emph{observation} denotes the physical interaction that establishes an experimental context, whereas \emph{measurement} refers to the logical determination and registration of a definite outcome within that context. In the classical regime (HCM), the two coincide operationally; in the quantum regime, contextuality decouples them.}

\subsection{Probability}
\label{prob}

Logical classicality, which physically entails the commutativity of preparable states and, consequently, the commutativity of physical observables, precludes the contextuality inherent to the quantum regime. In this manner, through Postulate~III, HCM posits a \emph{Kolmogorovian probability structure}, understood as \emph{epistemic information} concerning a physical reality that pre-exists observation and measurement. 

Kolmogorov's probability theory is founded on the following notions~\cite{Kolmogorov1950}:
\begin{itemize}
\item \emph{sample space} ($\Omega$), i.e., the set of all possible outcomes of an experiment;
\item \emph{event space} ($\mathcal{E}$), defined as a $\sigma$-algebra of subsets of $\Omega$;
\item \emph{probability measure} ($p$), i.e., a mapping $p \colon \mathcal{E} \to [0,1]$ satisfying the axioms of \emph{non-negativity}, \emph{normalization}, and \emph{additivity} over disjoint sets.
\end{itemize}
These concepts admit a direct correspondence within HCM:
\begin{itemize}
  \item the sample space corresponds to the set of classical states, $\{ \ket{\gamma_i} \bra{\gamma_i}\}$, which are mutually exclusive and collectively exhaustive;
  \item an event is represented by the experimental proposition $\mathfrak{p}_{\Delta}$, encoded by the propositional projector $\mathbb{E}_{\Delta}=\sum_{i \in \Delta} \ket{\gamma_i} \bra{\gamma_i}$;
  \item the probability of an event $\mathbb{E}_\Delta$, denoted as $p(\mathbb{E}_\Delta)$, is identified with its truth value $v_\rho(\mathfrak{p}_\Delta)$, as defined in~\eqref{truth_v}.
\end{itemize}
Explicitly, if the system is in the physical state $\rho = \sum_{i} p_i \ket{\gamma_i} \bra{\gamma_i}$, the probability is given by:
\begin{equation}
p(\mathbb{E}_\Delta) \doteq \operatorname{Tr}(\rho \mathbb{E}_\Delta) = \sum_{i \in \Delta} p_i.
\label{prob_class}
\end{equation}

Crucially, since all preparable states mutually commute, a \emph{joint probability} can be defined for the outcomes $a_k$ and $b_l$ of measurements of two observables $\hat{A}$ and $\hat{B}$, respectively. In fact, this quantity is well-defined in the Kolmogorovian scheme, and it is consistent with the cyclicity of the trace required in definition~\eqref{prob_class}:
\begin{equation}
  p(a_k,b_l) \doteq \operatorname{Tr}(\rho \mathbb{E}_{\Delta_k}^A \mathbb{E}_{\Delta_l}^B) = \sum_{i \in \Delta_k \cap \Delta_l} p_i, 
  \label{prob_cong}
\end{equation}
where $\mathbb{E}_{\Delta_k}^A$ and $\mathbb{E}_{\Delta_l}^B$ denote the corresponding measurement events. Therefore, for $p(b_l) > 0$, the conditional probability follows the standard Kolmogorov rule:
\begin{equation}
  p(a_k \mid b_l) = \frac{p(a_k,b_l)}{p(b_l)} = \frac{\sum_{i \in \Delta_k \cap \Delta_l} p_i}{\sum_{j \in \Delta_l} p_j}. \label{cond_prob}
\end{equation}
The ability to consistently define a joint probability for measurement events is the hallmark of a classical (and, in this context, epistemic) theory of probability. In this framework, the \emph{classical information}~\cite{Shannon1948} of the system in the state $\rho$ is defined as
\begin{equation}
I \doteq - K \operatorname{Tr}(\rho \ln \rho) = - K \sum_i p_i \ln p_i,
\label{info_c}
\end{equation}
where $K > 0$ is a suitable real constant. Since the measurement process in HCM is purely epistemic, $I$ quantifies the average ignorance about the exact classical state $\ket{\gamma_k} \bra{\gamma_k}$ that the measurement update resolves.

It is now possible to highlight the crucial distinction from the quantum model. Whereas in HCM the observational context is \emph{unique} and \emph{universal} -- and, consequently, a Hermitian operator that is not diagonal with respect to the unique classical basis is \emph{not} a physical observable -- the quantum framework admits a \emph{multiplicity} of observational contexts, each defined by a specific CSCO. Therefore, unlike in HCM, where a single sample space $\Omega$ underlies the $\sigma$-algebra of events, the possibility of mutually complementary contexts precludes the consistent definition of a universal joint probability. Thus, probability remains Kolmogorovian when restricted to a single fixed context, but the global structure arising from this coexistence is fundamentally \emph{non-Kolmogorovian}, characterized by interference and contextuality~\cite{Accardi1981}. In fact, for a pure superposition state $\rho = \ket{\psi}\bra{\psi}$, with $\ket{\psi} = \sum_i c_i \ket{\phi_i}$, the Born rule generates the interference correction strictly dependent on the terms $c_i^* c_j$ (with $i \neq j$):
\begin{equation}
    p(\mathbb{E}_\Delta) = \operatorname{Tr}(\rho \mathbb{E}_\Delta) = \sum_i |c_i|^2 \operatorname{Tr}(\ket{\phi_i}\bra{\phi_i} \mathbb{E}_\Delta) + \sum_{i \neq j} c_i^* c_j \braket{\phi_i | \mathbb{E}_\Delta | \phi_j}.
    \label{cond_prob_qm}
\end{equation}

Within this framework, one may consider transformations between classical bases associated with \emph{complementary} contexts. A Hermitian operator that is \emph{not} diagonal with respect to the initial classical basis but diagonal with respect to the final one \emph{does} actually constitute a physical observable in the sense of Postulate~II. The existence of such a basis transformation underscores that the relevance of the non-Kolmogorovian probability associated with the multiplicity of possible observational contexts is \emph{not} merely epistemic, but \emph{physical}. Since the probability of an outcome depends intrinsically on the specific experimental arrangement (the choice of context) rather than on a pre-existing state of affairs, probability itself becomes a feature of the physical interaction between system and apparatus. Accordingly, \emph{quantum information}~\cite{BBCJPW1993} is quantified by
\begin{equation}
I \doteq - K \operatorname{Tr}(\rho \ln \rho),
\label{info_q}
\end{equation}
which is formally identical to the expression in~\eqref{info_c}. However, its interpretation is fundamentally distinct: due to the contextuality discussed above, it represents a physical resource, encoding the fundamental limits imposed by the non-classical correlations arising from the superposition of states. Thus, in contrast to HCM, the post-measurement state update prescribed by the L\"uders rule is no longer merely epistemic: \emph{it constitutes a genuine physical transformation of the state, reflecting the objective modification of the system induced by the measurement within a specific observational context}.

\subsection{Time evolution}
\label{tevol}

Postulate~IV, when constrained by classicality, yields a key consequence: \emph{in HCM, the state of a closed (isolated) system remains constant in time}. Indeed, the unitary evolution prescribed by the quantum formalism 
\begin{equation}
i \hbar \frac{d \rho}{dt} = [H, \rho],
\label{vNeq}
\end{equation}
becomes trivial within HCM, owing to the commutativity between the state $\rho$ and the Hamiltonian $H$ imposed by the condition of observability. This gives rise to the ``problem of time''~\cite{Thebault2021}, implying that the issue of the ``nature of time'' is inherent in the interplay between the quantum and classical formalisms.\footnote{The ``problem of time'' signals that canonical quantum gravity retains a classical ontology, enforcing timeless structures through commuting states. Rather than a failure of quantum theory, this highlights the need for an approach where time emerges from contextual, non-commutative relations, effectively dissolving the problem~\cite{Rovelli1991,Barbour1994}.} In other words, \emph{within the unique and fixed universal classical context, non-trivial unitary dynamics of the state is unobservable}; for a closed system, the only empirically accessible behavior is \emph{stationarity}. This stationarity should not be interpreted as a physical prediction of timelessness, but rather as the kinematic signature of a framework that deliberately isolates the logical-probabilistic skeleton of classicality, leaving the Hamiltonian flow to be specified externally via phase-space dynamics (as detailed in Section~\ref{phspcm}). Consequently, for a state $\rho = \sum_i p_i \ket{\gamma_i} \bra{\gamma_i}$, the probabilities $p_i$ are time-independent and the underlying dynamics is deterministic.\footnote{If the system were physically coupled to an external environment, its dynamics would generally be stochastic and governed by a Markovian master equation. In that scenario, stationarity corresponds to the detailed balance condition. However, such physical openness lies outside the scope of the HCM measurement framework.} How can one reconcile this circumstance with the fact that standard classical dynamics is not stationary at all? The answer to this question lies in noting that, within the standard classical regime, in the limit $\hbar \to 0$, Eq.~\eqref{vNeq} is automatically satisfied by any evolution of $\rho$, and it merely entails that the Hamiltonian is a physical observable regardless of the state dynamics. Therefore, Eq.~\eqref{vNeq} does not necessarily entail dynamical triviality; in particular, it admits the ordinary conservation of the statistical ensemble, while temporal evolution is encoded in the Liouville flow of the probability weights $p_k(t)$ along phase-space trajectories. However, within the HCM framework, the limit $\hbar \to 0$ is not invoked as a direct parameter substitution; classical dynamics instead emerges via a structural correspondence that replaces the Hilbert-space commutator with an external Poisson bracket, thereby separating the logical skeleton of HCM from the Hamiltonian flow.\footnote{A bare $\hbar\to 0$ limit in Eq.~\eqref{vNeq} yields a mathematical identity $0=0$; the classical trajectory requires the additional symplectic structure that HCM deliberately delegates to an external phase-space description.}

Consistent with the physical stationarity predicted by Eq.~\eqref{vNeq}, even the discontinuous state update prescribed by the L\"uders rule does not correspond to a physical evolution, but rather signals that the measured system is being treated as epistemically accessible during the observational act. This epistemic transition is encoded in the variation of classical information, given by Eq.~\eqref{info_c}, which characterizes the observer's knowledge update. Since measurement in HCM resolves the ignorance about the pre-existing classical state, the classical information vanishes once the measurement is finalized, and the pure epistemic meaning of the state transition implies the \emph{identification between observation and measurement in the classical regime}.\footnote{Formally, this corresponds to Bayesian conditioning on a single outcome of a pre-existing statistical ensemble, which is mathematically distinct from the non-linear, context-dependent quantum measurement channel.} Given that classical information is purely epistemic, the distinction between ``closed'' and ``open'' must not be interpreted as a physical feature of the system: \emph{it pertains exclusively to the observational interface and, more specifically, to the purely epistemic content of the measurement act itself}.

In stark contrast to the HCM framework, the quantum regime does not admit a unique, universal observational context. The possibility of mutually complementary contexts fundamentally alters the merely epistemic status of both state evolution and information. Prior to measurement, the prepared object undergoes unitary evolution as prescribed by Postulate~IV. If the observational context includes the Hamiltonian as a descriptive observable -- which we term \emph{essential} for the system (see Appendix~\ref{gloss}) -- and the system is prepared in an eigenstate of the Hamiltonian, the state evolution reduces to the stationarity of HCM. Otherwise, the unitary evolution is not trivial and, according to the condition of observability, \emph{it signifies a non-essential recontextualization} that is \emph{empirically detectable} via interference effects. This unitary flow strictly preserves the quantum information defined in~\eqref{info_q}, implying that it remains invariant under reversible state rotations. In this way, the superposition structure evolves deterministically without any alteration of its intrinsic \emph{physical} informational content until the observation is completed by a measurement, which constitutes the new preparation of the system's state. 

Since non-trivial unitary evolution is physical, even the discontinuous update dictated by the L\"uders rule cannot be reduced to a mere epistemic revision. Rather, as highlighted above, it refers to a genuine \emph{physical transition} between contexts resulting in the generation of quantum indetermination. This means that, contrary to the classical regime, \emph{in the quantum regime observation is distinguished from measurement}: only upon measurement completion does the system abandon its \emph{observed} superposition state and actualize a determinate \emph{measured} outcome within the new context. Thus, the contextual shift drives an \emph{informational determination} wherein the information is irreversibly imprinted on a physical record, thereby constituting the intrinsic ``arrow of time''.\footnote{The irreversibility here is structural and contextual rather than thermodynamic; a full account of the thermodynamic arrow would require coupling to an environment~\cite{Zurek2003,Zurek2009}, which lies outside the present idealized framework.} 

Based on the preceding discussion, while the measurement outcome is always objective -- i.e., unambiguous and logically communicable -- the observational outcome is objective solely in the classical regime (where measurement and observation coincide) and non-objective in the quantum regime (where such an identification ceases to hold). Therefore, objectivity corresponds to an exclusively epistemic meaning of the wave function and L\"uders jumps; non-objectivity, on the other hand, corresponds to a meaning that is \emph{not} exclusively epistemic for these concepts. 
The conceptual and philosophical constraint that L\"uders jumps must be \emph{exclusively and universally} physical, and not epistemic as in the case of the measuring apparatus in the Copenhagen interpretation, leads to the traditional ``problem of measurement''~\cite{Margenau1963}, another facet of the ``problem of time,'' indicating that a fully satisfactory ontology must reconcile contextual actualization with macroscopic definiteness.

\subsection{HCM and phase space}
\label{phspcm}

In standard classical mechanics, the state $\rho(t)$ of a particle can be represented in \emph{phase space} by encoding the observer's knowledge of its position $Q$ and momentum $P$ at time $t$. This constitutes a probability distribution over the $(Q,P)$ points, specified at a given instant $t$ (the \emph{snapshot representation}) and subsequently updated in time according to the dynamical process referred to as \emph{motion}. It is, however, equally valid to adopt the \emph{trajectory representation}, wherein the state of the system does not correspond to a ``snapshot'' but rather \emph{provides a complete specification of the particle's initial conditions for motion, together with an associated probability distribution}. Within this framework, the trajectory is not identified with the state itself; rather, it is implicitly defined by the state and the dynamical evolution law.

The trajectory representation of classical mechanics in phase space is useful for understanding the formalism of HCM and for resolving the ``stationarity paradox'' of the state discussed in Section~\ref{tevol}. In fact, the correspondence between the trajectory representation in phase space and the HCM description can be explicitly articulated through the following points.\footnote{The finite-dimensional premise implies a discretized phase-space partition; the continuum is recovered formally by refining the cell size while treating $\hbar$ as a fixed elementary phase-space volume.} 
\begin{itemize}
    \item A \emph{single trajectory} is associated with the projector $\mathbb{P}_k$, representing the initial condition $(Q_k^0, P_k^0)$ on an energy surface.
    \item The \emph{ensemble of possible trajectories} is associated with the state $\rho = \sum_k p_k \mathbb{P}_k$, representing a probability distribution over initial conditions.
    \item The \emph{equation of motion} is associated with the Hamiltonian $\hat{H} = \sum_k E_k \mathbb{P}_k$, which defines the trajectory and the probability flow through it.
    \item \emph{Temporal evolution} is associated with the map $p_k(0) \to p_k(t)$, describing the motion of the ``probability packet'' along predefined trajectories.
    \item The \emph{measurement} revealing the initial condition of the current trajectory is associated with the epistemic reduction of the state according to the L\"uders rule, thus representing the Bayesian updating of the observer's knowledge.
\end{itemize}
To ground this mapping, consider a discretized phase space where basis vectors $\{\ket{\gamma_k}\}$ represent elementary cells $(Q_k, P_k)$ and a pure state $\mathbb{P}_k$ denotes definite localization within the $k$-th cell. Temporal evolution is not generated by the Hilbert-space commutator, but is instead delegated to an external Liouville flow updating the cell index deterministically ($k \to k(t)$). Thus, measurement merely reveals the occupied cell, collapsing epistemic uncertainty ($p_k \to 1$) without physically perturbing the trajectory.

The correspondence outlined above frames the HCM model within the context of geometric classical mechanics, wherein the underlying algebraic structure is determined by mutually commuting diagonal operators representing states and observables, thereby defining the phase space as a ``space of possibilities''. This phase space constitutes a static and timeless entity, within which the observed system undergoes deterministic or stochastic evolution. Within this descriptive framework, the state encapsulates the entire history of the system, ranging from its probable current configurations to the full spectrum of possible past and future trajectories, all of which are encoded in the initial conditions and the governing dynamical law. The reduction of unitary evolution to the trivial identity case constitutes the formal signature of this completeness: once the state is specified, its evolution is defined by the possibly stochastic dynamics of probabilities. This represents the essence of Laplacian determinism~\cite{Earman1986} extended to the stochastic case. Such a description entails a deliberate methodological trade-off: HCM isolates the static logical-probabilistic skeleton of classicality by construction and treats the phase-space flow as an externally specified dynamical input, thereby prioritizing epistemic consistency over internal dynamical generation.\footnote{The internal generation of classical phase-space flow within a Hilbert space is achieved in the Koopman-von Neumann formalism~\cite{KoopmanvNeumann1932}, which implements Hamiltonian dynamics in the $\hbar \to 0$ regime via a self-adjoint Liouvillian; HCM and KvN are thus complementary, isolating, respectively, the logical-probabilistic skeleton and the dynamical generator of classical theory.}

The trajectory-based interpretation extends to composite systems through the structural correspondence between the tensor product of Hilbert spaces $\mathcal{H}_A \otimes \mathcal{H}_B$ and the Cartesian product of phase spaces $\Gamma_A \times \Gamma_B$. Classicalized Postulate~V formalizes this structure by restricting admissible states to convex combinations of product projectors -- i.e., to \emph{separable} labeled states:
\begin{equation}
\rho_{AB} = \sum_{i,j} p_{ij} \, \mathbb{P}_i^A \otimes \mathbb{P}_j^B,
\label{comp_state}
\end{equation}
thereby excluding coherent superpositions and ensuring that the composite system admits a description in terms of a joint probability distribution. Within this framework, local observables take the form $\hat{O}_A \otimes \mathbb{1}_B$ or $\mathbb{1}_A \otimes \hat{O}_B$, preserving the dynamical independence of the subsystems unless an explicit interaction term is introduced into the Hamiltonian. This mirrors the traditional picture, in which trajectories evolve independently in the absence of coupling forces. Note that Postulate~V also applies consistently in the case of a \emph{hybrid} universe, featuring a classical subsystem described by the HCM model and a quantum subsystem. Even in this hybrid case, entanglement among the parts is prevented. This is enforced by the global commutativity constraint, which axiomatically excludes any non-classical correlations (including hybrid entanglement or quantum discord) across the composite system. A significant example in this regard is the case of the measuring apparatus and the measured object viewed within the framework of the Copenhagen interpretation, wherein the former always necessitates an objective, and thus classical, wave function.

Although standard phase-space formulations typically treat subsystems as labeled, labeling is not a logical prerequisite for classicality, which is instead defined by the Boolean structure of propositions and state commutativity. Thus, Postulate~VI builds upon this insight to address indistinguishable particles by restricting the admissible state space to symmetric or antisymmetric subspaces. In the trajectory representation, this corresponds to replacing ordered tuples with unordered multisets, thereby addressing the Gibbs paradox through the elimination of artificial microstate overcounting~\cite{Saunders2018}. Crucially, this constraint does not reintroduce quantum contextuality or entanglement but rather yields a classical probability distribution over configurations of indistinguishable parts. In this way, the postulate provides a Hilbert-space justification for the combinatorial adjustments in classical statistical mechanics, ensuring thermodynamic consistency while strictly adhering to the classical ontology.

\section{Quantum completeness and HCM}
\label{hcmonto}

Any formulation of a physical theory presupposes scientific realism, which entails a certain degree of independence of physical reality from the act of observation. In the debate over the completeness of quantum theory, neither Einstein nor Bohr departs from this fundamental epistemological premise. As noted in Section~\ref{intro}, the divergence lies exclusively in the meaning of \emph{objectivity}: for the EPR authors, it consists in an existence that remains \emph{physically independent} of the choice among multiple experimental setups; for the Copenhagen School, it consists in an existence that is \emph{logically communicable} relative to a given experimental setup. These two distinct theoretical frameworks yield markedly divergent epistemological consequences: for the EPR authors, an \emph{equivalence} is posited between objectivity and physical reality; for the Copenhagen School, objectivity constitutes \emph{merely a sufficient condition} for physical reality. Therefore, they inevitably reach opposing conclusions regarding the completeness of quantum theory. Recalling Section~\ref{hcm}, the two divergent points of view can be outlined as follows:
\begin{itemize}
\item \emph{EPR authors:} $\text{objectivity} \Leftrightarrow \text{physical reality}$. 
    
Physical reality must be \emph{objective} by definition, and it is represented by an \emph{epistemic} wave function. 

When referring to physical manifestations of the object, \emph{observation} and \emph{measurement} are synonymous terms, and the description provided by the wave function is generally ontologically \emph{incomplete}. 
\item \emph{Copenhagen School:} $\text{objectivity} \Rightarrow \text{physical reality}$.
    
Physical reality can be:
\begin{itemize}
\item \emph{Objective} -- i.e., the \emph{phenomenon} -- when it consists of physical manifestations of the object that are both observed \emph{and} measured (e.g., detection spots in the double-slit experiment);\footnote{The observed existence of the measuring instrument is, by definition, always objective; it constitutes the phenomenological unity along with the objective existence -- both observed and measured -- of the object.} it is represented by an \emph{epistemic} wave function;
\item \emph{Non-objective} when it consists of physical manifestations of the object that are observed but \emph{not} measured (e.g., interference fringes in the double-slit experiment); it is represented by an \emph{ontic} wave function.
\end{itemize}

When referring to physical manifestations of the object, \emph{observation} and \emph{measurement} are \emph{not} synonymous terms, and the description provided by the wave function is ontologically \emph{complete}.  
\end{itemize}
The requirement that the wave function possess a purely epistemic status is operationally implemented by the EPR authors through the EPR criterion. Yet, within the Hilbert-space formalism and under the joint assumptions of no-signaling and measurement independence, this criterion necessarily entails -- as we shall show below -- the impossibility of reproducing the entire quantum phenomenology. Before proceeding, however, a closer examination of the ontology implicit in HCM is warranted.

\subsection{HCM ontology}
\label{cpr}

The HCM framework affords an epistemic interpretation of the physical state $\rho = \sum_i p_i \ket{\gamma_i}\bra{\gamma_i}$, once the universal observational context is specified via the appropriate system observables. By definition, measurement does not alter the physics of the system under observation; rather, it resolves the uncertainty concerning the physical reality that pre-exists the act of observation. Since, according to the eigenstate-eigenvalue link, the system is found with certainty (i.e., with probability equal to unity) in the state encoded by the projector $\mathbb{P}_k = \ket{\gamma_k}\bra{\gamma_k}$ following the measurement, the validity of the EPR criterion allows us to identify these formal constituents as the ``elements of physical reality'' that ground the ontology described by the theory. In other words, the EPR criterion is structurally valid as a \emph{theorem} within HCM, once the classical states are recognized as the ``elements of physical reality'' formalized by the model. Although the HCM framework is conceptually and formally grounded in the commutativity of preparable states, the extension of mutual commutativity to observables precludes an ontological distinction between the latter and their associated classical states. This yields the epistemological consequence highlighted above: the impossibility of distinguishing \emph{observation}, understood as the pure \emph{physical contextualization of the object}, from \emph{measurement}, i.e., its \emph{contextualized logical determination} and, therefore, \emph{objectification}.

As already noted, the distinction between observation and measurement echoes the dichotomy between reality and objectivity: the ``interference pattern'' in the double-slit experiment is the outcome of observation rather than measurement; precisely for this reason, it carries a connotation of non-objectivity, while nonetheless being real. By contrast, the ``single spot'' in the same experiment is the outcome of measurement, not merely of observation, and for this reason it is characterized by objectivity. Only the latter constitutes, in the proper sense and according to the Copenhagen School, the \emph{physical phenomenon}. However, this distinction finds no place within the strong scientific realism espoused by the EPR authors, nor is it required by the ontology of HCM. Indeed, equating observation with measurement reflects the equivalence between objectivity and reality, thereby establishing HCM as an ontologically \emph{complete} theory in which every ``element of physical reality'' finds a counterpart in the classical state, as required by the EPR condition.\footnote{Although structurally complete, the HCM theory is ``empirically incomplete'' with respect to quantum mechanics, as it does not reproduce, by construction, all the quantum effects observed in the atomic world.} 

Specifically, the ontological structure of HCM can be defined as a ``local, separable realism'' that follows from the conjunction of the postulate of classicality and the standard Hilbert-space framework. The formalism implements a \emph{value-definite realism}, wherein physical observables possess determinate values at all times, encoded in the diagonal elements of the physical state $\rho$. By Postulate~V, the Hilbert space of a composite system comprising subsystems $A$ and $B$ is the tensor product $\mathcal{H}_{AB} = \mathcal{H}_A \otimes \mathcal{H}_B$. However, the classicality postulate restricts the admissible states to convex combinations of product projectors, as explicitly stated in Eq.~\eqref{comp_state}. This structure \emph{excludes by construction} coherent superpositions and non-separable (entangled) states. Consequently, the joint system admits a decomposition into well-defined marginal states $\rho_A = \operatorname{Tr}_B(\rho_{AB})$ and $\rho_B = \operatorname{Tr}_A(\rho_{AB})$, each of which retains the form of a classical mixture.\footnote{As noted in Section~\ref{phspcm}, in the trajectory representation, this corresponds precisely to the Cartesian product of classical phase spaces $\Gamma_A \times \Gamma_B$, where the state of the composite system is fully specified by the joint probability distribution $\{p_{ij}\}$ over locally defined initial conditions.} The formalism thus satisfies the standard criterion of \emph{separability}: \emph{the state of a composite system supervenes entirely on the states of its constituents and their classical correlations}.

Locality in HCM follows directly from the separable state structure and the Kolmogorovian probability framework. Consider the experimental propositions $\mathfrak{p}_{\Delta_k}^A$ and $\mathfrak{p}_{\Delta_l}^B$, with their corresponding propositional projectors $\mathbb{E}_{\Delta_k}^A$ and $\mathbb{E}_{\Delta_l}^B$. Owing to their mutual commutativity, the joint probability for the measurement of local observables $\hat{O}_A$ and $\hat{O}_B$ is well-defined:
\begin{equation}
    p(a_k, b_l) = \operatorname{Tr}_{AB}\!\left[ \rho_{AB} \left( \mathbb{E}_{\Delta_k}^A \otimes \mathbb{E}_{\Delta_l}^B \right) \right] = \sum_{i \in \Delta_k} \sum_{j \in \Delta_l} p_{ij}.
\end{equation}
The marginal distribution for subsystem $B$ is obtained by summing over the outcomes of $A$:
\begin{equation}
    p(b_l) = \sum_{j \in \Delta_l} \left( \sum_i p_{ij} \right) \doteq \sum_{j \in \Delta_l} p_j^B,
\end{equation}
which depends exclusively on the initial joint preparation $\{p_{ij}\}$ and is manifestly independent of the choice of measurement settings $\{\Delta_k\}$ or the outcome obtained on $A$. Operationally, even after a local measurement on $A$, the conditional state $\rho_B^{(k)}$ reflects only an epistemic update regarding pre-established correlations. Without classical communication, no statistical signaling can be detected on $B$. Consequently, no superluminal influence or nonlocal state collapse occurs; physical effects propagate solely through local Hamiltonian couplings, while apparent nonlocal correlations are strictly epistemic and factorizable. Thus, the theory provides a realization of Einstein's principle of local causality, embedded directly within the algebraic structure of the state space.

\subsection{EPR argument and the HCM reduction}
\label{epr}

In arguing for the completion of quantum mechanics, the EPR authors posit that any complete physical theory must satisfy the following joint requirements:
\begin{enumerate}
\item It allows for \emph{measurement independence}, understood as the experimenter's free choice of measurement settings;
\item It upholds \emph{locality} to ensure compliance with the no-signaling principle, as mandated by special relativity~\cite{Howard1985};
\item Its underlying ontology satisfies the \emph{EPR criterion} of physical reality, which secures the strong realism expected of natural laws.
\end{enumerate}
The EPR authors raised no objections to the Hilbert-space quantum formalism introduced by von Neumann, at least within its predictive domain, although they did not accept that the physical description provided by the quantum state was ontologically complete. On the other hand, the Copenhagen School clearly raised no objections to Assumption~1; nor did it have grounds to reject Assumption~2, which ensured compliance with the postulates of special relativity. From Bohr's perspective, the weakness of the argument therefore necessarily lay in Assumption~3 and the proposed ``criterion of physical reality''. Indeed, as we proceed to show, when translated into the Hilbert-space formalism, this criterion reduces quantum mechanics to the HCM model, thereby functioning as the postulate of classicality introduced in Section~\ref{class}.

In the Hilbert-space framework established by postulates~I-VI, the EPR criterion states that a conditional probability of $1$ for obtaining the outcome $o_k$ upon measuring the observable $\hat{O}$ in the physical state $\rho$ implies a predefined value $v_\rho(\hat{O}) = o_k$, independently of any actual measurement on the system. Locality requires that, for a bipartite system composed of subsystems $A$ and $B$, the reduced state of $B$ remains invariant under any measurement on $A$, reflecting the absence of instantaneous physical influence:
\begin{equation}
\rho_B = \operatorname{Tr}_A \rho_{AB} = \operatorname{Tr}_A \rho_{AB}',
\label{rhoinv}
\end{equation}
where $\rho_{AB}'$ denotes the updated state prescribed by the L\"uders rule following the measurement on $A$. Finally, measurement independence entails that the choice of measurement settings on $A$ is statistically independent of the hidden variables $\lambda$ (or the complete state specification), the joint state $\rho_{AB}$, and the physical properties of subsystem $B$. Consider, therefore, a bipartite system prepared in a state $\rho_{AB}$ that exhibits perfect correlations. Choosing to measure the observable $\hat{O}_1^A$ on subsystem $A$ yields an outcome $o_1^A$, which, by virtue of these correlations, guarantees that a subsequent measurement of $\hat{O}_1^B$ on $B$ will yield the outcome $o_1^B$ with probability equal to $1$. Locality ensures that the measurement on $A$ does not disturb the spatially separated system $B$, while the EPR criterion allows us to infer that the value $o_1^B$ corresponds to a pre-existing ``element of physical reality''. Similar reasoning applies if one instead chooses to measure a different observable $\hat{O}_2^A$ on $A$. Since, by locality and measurement independence, the physical reality of $B$ cannot depend on the arbitrary choice of measurement setting on $A$, both values $o_1^B$ and $o_2^B$ must exist simultaneously as ``elements of physical reality'' pertaining to $B$.

Now, within the Hilbert-space formalism, an observable $\hat{O}$ for a system $S$ possesses a definite value $o$ in the state $\rho$ if and only if the associated statistical variance vanishes, i.e., $\rho$ constitutes a \emph{dispersion-free state}~\cite{Ballentine1970}:
\begin{equation}
\operatorname{Tr}(\rho \hat{O}^2) - [\operatorname{Tr}(\rho \hat{O})]^2 = 0.
\label{dispfree}
\end{equation}
Condition~\eqref{dispfree} is equivalent to $\rho$ being supported entirely on the eigenspace of $\hat{O}$ associated with $o$. The EPR criterion, together with locality and measurement independence, entails that \emph{any} pair of observables on $B$ that can be perfectly predicted via local measurements on $A$ must simultaneously correspond to pre-existing elements of reality. If one demands that the quantum state $\rho$ itself provides this complete description (i.e., that quantum mechanics is ontologically complete in the EPR sense), then $\rho$ would have to be simultaneously dispersion-free for non-commuting observables. However, standard Hilbert-space spectral analysis shows that no density operator can satisfy this condition unless the observables commute. Since, for \emph{maximally} entangled states, \emph{quantum steering}~\cite{JWD2007,UCNG2020} ensures that \emph{any} local observable on $B$ can be perfectly predicted by a suitable measurement choice on $A$,\footnote{In a maximally entangled state, the Schmidt decomposition guarantees perfect correlations in any basis. For every observable $\hat{O}_B$ on $B$, there exists an observable $\hat{O}_A$ on $A$ such that a measurement of $\hat{O}_A$ predicts the outcome of $\hat{O}_B$ with probability $1$. Consequently, the EPR criterion applies to the entire algebra of local observables on $B$.} the \emph{universal} applicability of the EPR criterion forces \emph{the algebra of local observables on $B$ to be commutative}. This mutual commutativity entails that all measurement observables belong to a single observational context and share a common eigenbasis. According to the condition of observability stated in Section~\ref{stob}, this basis must constitute the context-defining basis and simultaneously diagonalize all preparable states, thereby corresponding to the classical basis. Thus, the lattice of experimental propositions becomes distributive, and the resulting phenomenology is non-quantum and governed by classical Boolean logic. This gives rise to a \emph{trilemma} whereby accepting the Hilbert-space formalism of quantum mechanics necessitates abandoning at least one of the following: measurement independence (experimental free choice), locality, or strong realism (the EPR criterion of physical reality). 

It is instructive to emphasize how quantum steering functions as a conceptual link between quantum and classical phenomenology, mediated by the EPR criterion. In the absence of the EPR criterion, maximally entangled states represent the most striking manifestation of the ontic nature of the quantum state~\cite{PuseyBarrettRudolph2012}. Once the EPR criterion is assumed, however, these states instead serve as a formal bridge to the classical framework. Within this resulting non-contextual structure, they no longer correspond to physical states \emph{per se}, but rather to formal entities that are \emph{empirically indistinguishable from uniform statistical mixtures}, thereby signaling a decisive shift toward a purely epistemic characterization of the state.\footnote{Recall the role of maximum-entropy states in formalizing the postulate of classicality, as expressed in Eq.~\eqref{prep_s} of Section~\ref{class}. The empirical indistinguishability of maximally entangled states from uniform statistical mixtures resonates with Einstein's statistical interpretation of quantum mechanics, wherein the quantum state encodes ensemble properties rather than individual realities.} In this way, the resulting theory discards any non-objective ontology; for the EPR authors, this means making quantum theory ontologically complete. Yet the criterion was introduced \emph{axiomatically}, by presupposing an ontology \emph{assumed} to be fully compatible with quantum phenomena; instead, a genuine criterion of physical reality can only be \emph{derived} from ontology. In fact, the EPR criterion effectively operates in the theory as a ``postulate of classicality,'' and such a criterion can only be derived from a \emph{classical} ontology. For this reason, from the standpoint of the Copenhagen School, the elimination of non-objective phenomenology by means of the EPR criterion not only fails to constitute a completion of quantum theory but, on the contrary, renders the theory ontologically incomplete, as it suppresses the description of real physical effects that are not unambiguously communicable and are ontologically distinct from those that are.

\subsection{Bell's argument}
\label{bell}

Imposing the EPR criterion on the Hilbert-space quantum formalism yields a deterministic theory that \emph{renders empirically inaccessible} the coherence contributions associated with the off-diagonal elements of the system's state. This theory is precisely the HCM model, in which the ontic constituent corresponds to a definite projector $\mathbb{P}_i = \ket{\gamma_i}\bra{\gamma_i}$, while the probabilities $p_i$ encode the observer's ignorance. The label $i$ plays the role of the model's ``hidden variable'' -- i.e., the underlying ``element of physical reality'' posited by the theory -- and incomplete knowledge of this variable is represented by a Kolmogorovian $p_i$, which fully characterizes the system's statistical state. This raises a key foundational question: does there exist a deterministic theory that posits hidden variables $\lambda$, analogous to $i$ in the HCM, governed by a Kolmogorovian probability $p_\lambda$, yet capable of \emph{incorporating} the empirical quantum effects associated with the off-diagonal terms of the density matrix? 

This is, in fact, a long-standing issue that has permeated the history of quantum foundations since Born's seminal work on the probabilistic interpretation of the theory~\cite{Born1926}. It was widely regarded as settled by von Neumann in his 1932 treatise~\cite{vNeumann1955}, at least until the criticisms raised by Bell in 1966~\cite{Bell1966}.\footnote{The question gave rise, beginning with Bohm's seminal work~\cite{Bohm1952}, to the interpretative research program in quantum theory known as ``hidden-variable theories'', which continues to represent one of the alternatives to the Copenhagen interpretation.} The originality of Bell's approach to this issue lies in its focus on probability, through the introduction of the notion of \emph{local causality}~\cite{Bell1981} -- independent of the specific formalism of the theory -- thereby enabling an experimental approach to the problem of quantum ontology~\cite{Aspect1976,Lamehi-RachtiMittig1976}. In fact, local causality captures the physical principle of ``no superluminal influence'' through the statistical independence condition for the measurement probability $p_\lambda$:
\begin{equation}
p_\lambda (\mathcal{A},\mathcal{B}|a,b) = p_\lambda(\mathcal{A}|a) p_\lambda (\mathcal{B}|b),
\label{lcaus}
\end{equation}
where $\mathcal{A}$ and $\mathcal{B}$ denote the measurement outcomes on the two subsystems $A$ and $B$, respectively, $a$ and $b$ denote the corresponding measurement settings, and $\lambda$ denotes a complete state specification (which may encompass hidden variables or, in principle, the quantum state itself should the theory be complete). 

On the other hand, the local causality expressed by~\eqref{lcaus} is equivalent to the conjunction of two independent conditions~\cite{Jarrett1984,Shimony1984}:
\begin{itemize}
  \item \emph{parameter independence}, i.e., $p_\lambda (\mathcal{B}|a,b) = p_\lambda(\mathcal{B}|b)$ for all $a$; 
  \item \emph{outcome independence}, i.e., $p_\lambda(\mathcal{A},\mathcal{B}|a,b) = p_\lambda(\mathcal{A}|a,b) p_\lambda(\mathcal{B}|a,b)$.
\end{itemize}
These are in direct correspondence with the ontological assumptions of the EPR argument in Section~\ref{epr}. Specifically:
\begin{itemize}
  \item \emph{Locality} is mapped, at the probabilistic level, onto \emph{parameter independence} and translates, within the Hilbert-space formalism, into the invariance condition for $\rho_B$ expressed by Eq.~\eqref{rhoinv}; should the condition of parameter independence be violated, the choice of measurement setting $a$ would statistically influence the outcomes on $B$ even for a fixed $\lambda$, thereby implying action at a distance.\footnote{Parameter independence is a stronger condition than \emph{operational} no-signaling. Bohmian mechanics provides a paradigmatic example: it explicitly violates parameter independence. Nevertheless, the theory respects no-signaling at the statistical level due to the \emph{quantum equilibrium hypothesis} ($\rho = |\psi|^2$), which ensures that the marginal distribution of outcomes on $B$ remains independent of the measurement setting chosen on $A$.}
  \item The \emph{EPR criterion} is mapped, at the probabilistic level, onto \emph{outcome independence} and translates, within the Hilbert-space formalism, into the requirement of a dispersion-free state as expressed by Eq.~\eqref{dispfree}; should the condition of outcome independence fail, the measurement outcome on $A$ would probabilistically influence the outcomes on $B$ (or vice versa) even for a fixed $\lambda$, thereby contradicting the assumption that $\lambda$ provides a complete specification of the physical state.\footnote{Since no $\lambda$ can simultaneously assign definite values to incompatible observables while preserving the factorization of conditional probabilities, the EPR criterion cannot be upheld.}
\end{itemize}

Building on the ontological-probabilistic correspondence outlined above, the trilemma highlighted in Section~\ref{epr} maps directly to the joint requirements of measurement independence, parameter independence, and outcome independence, which are formally equivalent to Bell's dilemma between measurement independence and local causality~\cite{Bell1964}. Within the Hilbert-space formalism, the satisfaction of the Bell inequalities guarantees the existence of a global joint probability distribution, which, for projective measurements, implies the mutual commutativity of the corresponding observables. This condition, in turn, reduces the theoretical structure precisely to the HCM model: the lattice of experimental propositions becomes distributive, statistical states assume a Kolmogorovian form, and the framework exhibits the logical classicality that characterizes the HCM. Consequently, satisfaction of the Bell inequalities is tantamount to endorsing the HCM ontology. This structural reduction is captured by Fine's theorem~\cite{Fine1982}, which establishes that the satisfaction of the Bell inequalities is necessary and sufficient for the existence of a global joint probability distribution. In the bipartite setting, this requirement renders local causality equivalent to \emph{spatial non-contextuality} (cf. Appendix~\ref{gloss}): the marginal statistics at each subsystem become independent of the measurement context fixed at the remote site, thereby recovering the non-contextual, Kolmogorovian backbone of the HCM. The empirical violation of the Bell inequalities thus signals the breakdown of this joint distribution, forcing the framework into a genuinely contextual regime where the classical HCM structure can no longer be sustained. However, in the next section, we use the HCM model as a control baseline. By systematically relaxing the postulate of classicality while preserving the underlying Hilbert-space structure, we can isolate the specific ontological features that emerge when multi-contextuality is restored. This leads us to the proposal of a genuine quantum ontology.

\section{Quantum ontology}
\label{quonto}

The HCM model, described in Section~\ref{hcm} and employed in Section~\ref{hcmonto} as a control model for interpreting the EPR argument concerning the ontological completeness of quantum theory, \emph{does not} represent a classical theory in the conventional sense. Rather, it constitutes an epistemological classicality \`a la Bohr, reflecting the observer's Boolean logic without imposing prior ontological assumptions on the formalism. The ontology inherent to the physical model can be inferred \emph{a posteriori}, once the \emph{epistemological} requirement of classicality has been implemented within the Hilbert-space quantum framework. Consequently, the model does not arise as a limiting theory as $\hbar \to 0$, nor does it preclude identical particles, in contrast to standard classical ontology. Instead, it characterizes classicality as a \emph{single-context quantum theory}, i.e., a framework that retains the full Hilbert-space structure and Born rule while restricting physically admissible states and observables to a single, context-independent commutative subalgebra. In this sense, the uniqueness and universality of the observational context do not conflict with quantum multi-contextuality; rather, they provide the fixed reference framework relative to which alternative contexts are mathematically defined and operationally compared.

Indeed, HCM formalizes Bohr's requirement that the measuring apparatus must be described in classical terms: not as a macroscopic or $\hbar \to 0$ limit, but as a necessary epistemic condition for defining experimental arrangements and ensuring unambiguous communication of results~\cite{Bohr1928}. Consequently, HCM describes the apparatus and, more broadly, the \emph{environment} in which the system is embedded prior to measurement. The single-context quantum theory represented by HCM does not conflict with the standard classical formalism as long as the quantum of action is negligible relative to the characteristic action scales of the environment. More importantly, it emerges as the genuine epistemological classical theory \emph{even when these environmental actions remain of the order of $\hbar$, regardless of whether they are subsequently amplified and recorded}. This implies that, during the measurement interaction, the physical apparatus may possess action scales comparable to $\hbar$, just as the system does.\footnote{This position aligns with the view that Bohr's ``classical terms'' requirement is epistemological rather than thermodynamic or macroscopic. The apparatus need not be ``large'' in a naive sense; it must only instantiate a fixed Boolean context for unambiguous communication.} However, unlike the system, it is described within the non-contextual framework of HCM, thereby providing the fixed observational context that Bohr identified as the prerequisite for any objective physical statement.

\subsection{Single-context ontology}
\label{stobonto}

In Section~\ref{cpr}, we described the main features of the ontology underlying the HCM model, identifying the fundamental ontological elements exclusively with the classical states $\mathbb{P}_k = \ket{\gamma_k}\bra{\gamma_k}$. We now reframe the HCM ontology not as a ``universal classical'' framework, but as a single-context quantum ontology situated within the broader ontology of contextual physical phenomena. The formal elements to which ontological significance should be assigned are clearly the physical states and the observables, as introduced by Postulates~I and~II, respectively, in Section~\ref{hcm}.

We first emphasize that, given the epistemic meaning of the probabilities $p_i$ in $\rho=\sum_i p_i \mathbb{P}_i$, which reflect mere statistical ignorance, the formal elements that are genuinely \emph{ontically} distinct are exclusively represented by the classical states $\mathbb{P}_j$ and $\mathbb{P}_k$, for $j\neq k$. These correspond to an objective reality both in the sense of the EPR authors, by definition, and in the sense of the Copenhagen School, in that they correspond to the \emph{certain} experimental propositions formulated in~\eqref{exprop} -- i.e., those with probability one and vanishing statistical variance as given by Eq.~\eqref{dispfree}. We further note the specification of the measurement outcome in such propositions, thereby indicating that, within the single-context framework of HCM, only the realization of the measured value carries genuine epistemic significance, as formalized by the truth-value assignment in Eq.~\eqref{truth_v}. Consequently, the term ``observation'' must be understood as ``observation of the measurement outcome''. In this sense, observation is operationally equivalent to the measurement of a definite outcome. To avoid the category error of conflating the physical entity with the act of interrogating it, in what follows, we introduce a refined ontological vocabulary (see Appendix~\ref{gloss}). This terminology does not alter the mathematical formalism, but it prevents the conceptual flattening typical of standard treatments.

We now focus on the formal distinction between the state and the observable, noting that this also corresponds to a genuine ontological distinction, albeit not an ontic one of the sort that characterizes states themselves. To grasp the significance of this difference, one must examine the distinct implications of the property of \emph{commutativity}. Within the HCM formalism, as derived in Section~\ref{hcm}, the commutativity of operators manifests across three conceptually distinct levels, which we proceed to highlight:
\begin{enumerate}
\item commutativity among classical states, directly linked to classicality and the consequent epistemological emphasis;
\item commutativity between classical states and observables, which provides the necessary formal link -- i.e., the eigenstate-eigenvalue link -- between the domain of states and the domain of observables;
\item commutativity among observables, directly linked to their simultaneous physical measurability, irrespective of the specific measurement procedure employed.
\end{enumerate}

As previously noted, the mutual commutativity of classical states is related to their ontic distinction (see Appendix~\ref{gloss}). This commutativity permits the existence of a classical basis $\{\ket{\gamma_j}\}$ whose \emph{vectors} $\ket{\gamma_j}$ define the \emph{substantial} elements of the ontology: for $j \neq k$, two classical states $\mathbb{P}_j = \ket{\gamma_j}\bra{\gamma_j}$ and $\mathbb{P}_k = \ket{\gamma_k}\bra{\gamma_k}$ are ontically distinct because they correspond to distinct substantial elements $\ket{\gamma_j}$ and $\ket{\gamma_k}$. Likewise, the commutativity between a classical state $\mathbb{P}_j$ and an observable $\hat{O}$ corresponds to an ontological, though \emph{not} ontic, distinction. In fact, this commutativity implies that the classical basis can also be viewed as an eigenbasis of the observable. The commutativity between $\mathbb{P}_j$ and $\hat{O}$ is grounded in the \emph{common} substantial element $\ket{\gamma_j}$ -- via the eigenstate-eigenvalue link; therefore, $\mathbb{P}_j$ and $\hat{O}$ are not ontically distinct. In this sense, the observable $\hat{O}$ defines an \emph{existential} element of the ontology and fundamentally represents a \emph{physical manifestation} of the substantial element $\ket{\gamma_j}$. This separation is methodologically essential. Standard introductory treatments of quantum mechanics often identify normalized Hilbert-space vectors directly with physical states, a practice that tends to obscure the ontological status of classical basis vectors as substantial elements. Von Neumann's operator-theoretic formulation, by contrast, explicitly preserves this ontological distinction by treating states and observables as distinct classes of operators. 

Building on this foundation, the ontological distinction between a classical state, $\mathbb{P}_j$, and an observable, $\hat{O}$, is here termed the \emph{processional distinction} (see Appendix~\ref{gloss}), which stands alongside the \emph{ontic distinction} between distinct classical states $\mathbb{P}_j$ and $\mathbb{P}_k$. In this way, the processional distinction may be linked to the traditional philosophical distinction between ``substance'' and ``existence'' in a physical object~\cite{Kenny2006,FernandezMoujan2024}. The \emph{processional unity} that indissolubly binds a substantial element and an existential element, albeit in an ontologically distinguishable manner, without introducing new physical degrees of freedom, may be termed the \emph{character} of the system.\footnote{This conceptual partition formalizes the operational distinction between the set of physically preparable configurations (i.e., the state space) and the set of admissible interrogations (i.e., the observable algebra) within a fixed experimental context.} Furthermore, by analogy with the previous case, since the mutual commutativity of observables follows directly from the commutativity between classical states and observables, the distinction among observables likewise formally expresses a processional distinction. Indeed, the substantial element is shared as a common eigenstate $\ket{\gamma_j}$, whereas the existential elements, represented by different observables, are ontologically distinct.

Therefore, by recognizing the different formal origins of the various commutativity relations, one can discern their respective ontological significance. Failing to do so, treating every Hermitian operator merely as ``governed by spectral properties and commutation relations'' conceals the genuine ontological distinctions among them, thereby obscuring the difference between the substantial and existential levels, as well as that between the ontic and epistemic domains. We conclude the description of the single-context ontology of the HCM model by noting that, according to Eq.~\eqref{exp_v}, a formal change of basis in the theory's Hilbert space carries no empirical consequences -- i.e., it is not detectable through the measurement of any observable. This implies that the change at the substantial level is merely formal, while its existential import is vacuous. In other words, within the HCM framework, a basis transformation in the Hilbert space of states does not alter the processional unity that defines the system's physical character, and therefore entails no observable physical consequences.\footnote{From a foundational standpoint, HCM can be viewed as a formalization of the ``single-framework'' or ``single-history'' approach to quantum ontology, sharing structural affinities with Consistent Histories~\cite{Griffiths2002} and Modal Interpretations~\cite{Lombardi2026}, yet differing in that it derives classicality axiomatically from state commutativity rather than from decoherence or consistency conditions. This makes HCM well-suited as a control model for isolating the logical prerequisites of objectivity.} Having clarified the ontological status of the non-contextual (classical) regime, we now turn to the \emph{contextual} quantum ontology. By relaxing the postulate of classicality while preserving the Hilbert-space structure, we recover the multiplicity of experimental arrangements discussed in Section~\ref{prob}.

\subsection{Ontology of observation}
\label{obsont}

In Section~\ref{stobonto}, we identified two types of ontological distinction within a single observational context -- i.e., with respect to a fixed classical basis: an ontic distinction between distinct classical states, and a processional distinction between a classical state and an observable, or between two distinct observables. We further noted that a change of basis in the Hilbert space of states, provided it remains confined to a single observational context, bears no existential implications, leaving the character of the system unaltered. But what occurs when a change of basis at the substantial level entails a shift in the observational context? 

First, we note that during a contextual shift, only physical states admit meaningful ontological comparison across contexts. Comparisons between states and observables, or between observables of incompatible contexts, are ontologically ill-posed: they either conflate distinct operational categories (preparation vs. observation) or violate the non-commutative structure that precludes a joint ontological attribution. States, however, can be consistently tracked as they undergo unitary recontextualization, preserving their character as physically prepared systems. Furthermore, in general, \emph{any reversible physical transformation that implements a shift between mutually complementary observational contexts defines the phase of quantum observation}. Let us therefore assume that the physical state $\mathbb{P}_j = \ket{\gamma_j}\bra{\gamma_j}$ is prepared within a given context, and that a contextual shift is physically enacted by a unitary operator $U$ arising from the system-environment interaction. The substantial element identified by $\ket{\gamma_j}$ \emph{does not} change under this contextual shift: $\ket{\gamma_j}$ remains the same vector, albeit expressed in a new physical basis. Consequently, as expected, the change of context \emph{does not} entail an ontic change in the observed system. However, the ontological distinction between the initially prepared state $\mathbb{P}_j$ and the contextual projectors $\{\mathbb{P}_k'\}$ defining the final frame \emph{is not} processional, as $[\mathbb{P}_j, \mathbb{P}_k'] \neq 0$ in general. A third type of ontological distinction is therefore required, alongside the ontic and processional distinctions, to account for this novel situation. We designate this category as the \emph{tropos-existential distinction} (see Appendix~\ref{gloss}). Accordingly, the system in the initially prepared state $\mathbb{P}_j$ is in \emph{tropos-existential actuality} relative to its preparation context, whereas it is in \emph{tropos-existential potentiality} relative to the shifted context, reflecting the inherent indeterminacy of outcomes in the new observational frame. 

Likewise, temporal evolution, as prescribed by Postulate~IV (Section~\ref{hcm}) for closed systems, constitutes a change of physical basis, 
\begin{equation}
\rho(t) = U(t) \, \rho(0) \, U^\dagger(t),
\label{unievol}
\end{equation}
which can be subsumed under the observation phase. This corresponds to a unitary change of basis over time, governed by the system's Hamiltonian, and follows the preparation phase. During unitary observation, its character retains the general existential tropos of potentiality without undergoing any substantial change. The unitary evolution accompanying quantum observation is, nevertheless, characterized by a \emph{physical interaction} between the observational environment and the observed object. The closed, evolving universe is therefore \emph{always bipartite} and subject to Postulate~V. Both poles are necessarily required for observation and, in particular, for temporal evolution. How is the ontological distinction between them characterized? By definition, the character of the observational environment is prepared and stationary throughout the evolution: it exhibits an existential tropos of actuality, which is necessary for the logical communicability of the observation -- i.e., for its objectivity. In other words, it constitutes a single-context frame. Conversely, the observed object is generally non-stationary: it exhibits an existential tropos of potentiality that is unsuited to the logical communicability of observation, and its description is non-objective. In other words, it represents a multi-context frame.

Formally, the observational environment is described by a Hilbert space $\mathcal{H}_\text{env}$, and its state during the observation is given by $\rho_\text{env} = \sum_i p_i \ket{\epsilon_i} \bra{\epsilon_i}$, where $\ket{\epsilon_i}$ denotes, as usual, a classical-basis vector in $\mathcal{H}_\text{env}$ that defines the context.\footnote{If the observational environment is constituted by the experimental apparatus, the states of the classical basis $\ket{\epsilon_i}$ are the ``pointer states''.} Conversely, the observed object is described by a Hilbert space $\mathcal{H}_\text{obj}$, and its state during the observation takes the form $\rho_\text{obj} = \sum_i p_i \rho^{(i)}$, where $\rho^{(i)}$ is a conditional density matrix that evolves unitarily under the Hamiltonian $\hat{H}_\text{obj}$. Accordingly, the observational universe is described by a state $\rho$ defined on the composite Hilbert space $\mathcal{H} = \mathcal{H}_\text{env} \otimes \mathcal{H}_\text{obj}$, such that the reduced states satisfy $\rho_\text{env} = \operatorname{Tr}_\text{obj} (\rho)$ and $\rho_\text{obj} = \operatorname{Tr}_\text{env} (\rho)$:
\begin{equation}
\rho(t) = \sum_i p_i \ket{\epsilon_i} \bra{\epsilon_i} \otimes \rho^{(i)}(t).
\label{cpl_state}
\end{equation} 
To ensure that the environment acts as a stable, classical control register (fixing the context) while allowing the object to evolve conditionally, the interaction must not entangle the environment's classical-basis states. Therefore, the observational Hamiltonian $\hat{H}_\text{obs}$, as a Hermitian operator on $\mathcal{H}$, takes the form
\begin{equation}
\hat{H}_\text{obs} = \sum_i \ket{\epsilon_i}\bra{\epsilon_i} \otimes \hat{V}^{(i)}, 
\label{hobs}
\end{equation}
where $\hat{V}^{(i)}$ is a suitable potential defined on $\mathcal{H}_\text{obj}$ that characterizes the observation of the object~\cite{Zurek2003,Zurek2009}. \emph{Only Hamiltonians of the form given in~\eqref{hobs} qualify as valid observational Hamiltonians to describe the physical interaction between the observational environment and the observed object}. The specific structure of $\hat{H}_\text{obs}$ ensures that, while the entire observational universe is closed and evolves unitarily, the observed object undergoes conditional unitary evolution within each environmental branch labeled by $i$.\footnote{In the setup defined by Postulates~I-VI, the observational environment does not act as a dissipative thermal bath, but rather as a stable quantum control register that fixes the observational context.} Thus, \emph{decoherence constitutes the necessary physical outcome that actualizes the logical constraint of classicality}, formally implemented by means of the HCM framework for the observational environment. The conceptual distinction between the observational environment and the observed object thereby defines the \emph{Heisenberg cut}, which is formalized by the tensor-product decomposition $\mathcal{H} = \mathcal{H}_\text{env} \otimes \mathcal{H}_\text{obj}$. Although inherently movable, the cut is conventionally held fixed in standard experimental protocols; nevertheless, its mobility can be directly probed in decoherence experiments that explore the quantum-classical boundary~\cite{MKTKIWSMW2000,HUHRBZA2003}. Such experiments explicitly manifest the \emph{physical bootstrapping} mechanism governing the interplay between the observational environment and the observed system, thereby reflecting the \emph{logical} objectification underlying the transition between quantum and classical descriptions.

We conclude our account of the ontological aspects of the observation process by underscoring the connection between temporal evolution and information -- i.e., the \emph{informational evolution}. This evolution bears significance solely with respect to the observed system, 
\begin{equation}
I_\text{obj} (t) = - K \operatorname{Tr} [\rho_\text{obj} (t) \ln \rho_\text{obj}(t)],
\end{equation}
and not with respect to the global observational complex (which undergoes unitary evolution) or to the observational environment (which, insofar as it is classical, remains stationary). It generally constitutes an evolution of a tropos-existential character, grounded in the epistemic status ascribed to the environmental state in addition to the structural non-commutativity between contexts (as described in Section~\ref{prob}). In fact, the uncertainty of the environmental state causes the information associated with the reduced state of the observed object to vary over time, reflecting a tropos-existential evolution, whereas its certainty stabilizes this information, albeit without necessarily yielding knowledge thereof.

\subsection{Ontology of measurement}
\label{meaonto}

While observation is formally characterized by a unitary basis transformation in the Hilbert space $\mathcal{H}$ that, over time, evolves the initially prepared state of the observational universe without altering its ontic status, but only its tropos-existential character, the situation acquires a distinct ontological significance upon measurement. \emph{Measurement completion} (see Appendix~\ref{gloss}), indeed, is analyzed as comprising two conceptually distinct events within the bipartite observational universe that jointly constitute the objectification of the observed reality -- i.e., the physical phenomenon:
\begin{itemize}
\item the \emph{epistemic revelation} of the \emph{actual} (tropos-existential) character of the environment;
\item the \emph{ontic determination} of the \emph{potential} (tropos-existential) character of the object.
\end{itemize}

The objectification of the observed reality through measurement is formally expressed by the L\"uders rule prescribed by Postulate~III, both in the epistemic (single-context) formulation of the HCM model applied to the description of the environment, and in the ontic (multi-context) formulation of the quantum model applied to the description of the object:\footnote{The origin of the ``problem of measurement'' is now further clarified; in fact, it arises from presupposing a universal ontic meaning for the physical state of the observational universe, disregarding its bipartite distinction and dual tropos-existential condition.} 
\begin{equation}
\rho \rightarrow \frac{(\mathbb{P}_j^\text{env} \otimes \mathbb{P}_k^\text{obj}) \rho \, (\mathbb{P}_j^\text{env} \otimes \mathbb{P}_k^\text{obj}) }{\operatorname{Tr}[\rho \, (\mathbb{P}_j^\text{env} \otimes \mathbb{P}_k^\text{obj})]}.
\label{luders}
\end{equation}
However, unlike observation, the L\"uders rule prescribes a \emph{probabilistic} transformation of the composite physical state, with probability $\operatorname{Tr} [\rho (\mathbb{P}_j^\text{env} \otimes \mathbb{P}_k^\text{obj})]$, that \emph{cannot be reduced to a unitary basis transformation in $\mathcal{H}_\text{env}\otimes \mathcal{H}_\text{obj}$}. Indeed, the substantial element, $\ket{\epsilon_j} \in \mathcal{H}_\text{env}$, corresponding to the prepared state of the environment, $\rho_\text{env}$, remains invariant under the measurement transformation expressed by~\eqref{luders}, whereas the substantial element in $\mathcal{H}_\text{obj}$, corresponding to the prepared state of the object, $\rho_\text{obj}(0)$, is generally altered by it. This specific circumstance corresponds to two distinct ontological effects: \emph{the general ontic and tropos-existential transition of the measured character from potentiality to actuality}. 

From the epistemic characterization of the observational environment and its correlations with the object, one may reconstruct the specific unitary evolution of the system's state and, consequently, infer its initial state and its substantial element. This constitutes the preliminary phase of objectification, which extracts quantum information about the observed object without, in general, fixing its existential tropos. Such objectification is feasible insofar as the \emph{measuring apparatus} participates in the tropos-existential actuality of the observational environment. Put differently, it shares the environment's classical logical structure and, therefore, its operational accessibility. 

The second event of objectification, which ultimately constitutes the physical phenomenon and brings the measurement to completion, concerns the tropos-existential determination of the object's character by the same apparatus -- i.e., within the Hilbert-space formalism, the reduction of the system's wave function. In this event, the substantial element of the initially prepared state is typically replaced by a \emph{new}, contextually reconstituted substantial element. This ontic change thus corresponds to the actualization of the existential tropos of the measured character. 

The foundational question is: why is this second event inherently probabilistic and indeterministic? The answer lies in the nature of determinism, which requires a \emph{prior} state of definite properties (tropos-existential objectivity). The observational environment and the apparatus already possess this objectivity. However, the observed object, being in a state of tropos-existential potentiality, \emph{lacks} these definite properties prior to the measurement act. Since the measurement is precisely the process tasked with \emph{conferring} this objectivity upon the object, the outcome cannot be predetermined; it is actualized only at the moment of completion. Crucially, this demarcation precludes conflating the apparatus with the observed object, as they pertain to distinct logical regimes: the apparatus is classical and strictly observable, whereas the object is quantum and amenable to measurement. Upon completion of the measurement process by the measuring apparatus, and in accordance with the \emph{logos of measurement}, the existence of the measured object is rendered \emph{immediately actual}, and \emph{knowledge} thereof becomes classically shareable -- i.e., logically communicable -- as the measurement outcome among all observers operating within the same \emph{logos}.\footnote{This shared \emph{logos} addresses the paradox of quantum universalism highlighted by Frauchiger and Renner \cite{FrauchigerRenner2018}, showing that contradictions arise only when apparatuses and observers are treated as purely quantum systems without a common classical logical framework. Such a framework precisely enables a consistently shared tropos-existential actualization, thereby circumventing the theorem's no-go result.}

In contrast to mere observation, characterized by universal unitary evolution, the completion of measurement generally induces a \emph{discontinuous} change in the information content of the observed universe (comprising the observational environment and the observed object). This change corresponds to the ontic assignment of a new contextual substantial element and, consequently, to the tropos of actuality conferred upon the measured system. Such a variation reflects an informational evolution that possesses a specifically substantial significance, rather than a merely tropos-existential one as in the case of observation. The nature of this temporal flow is universal, nonlocal, and discontinuous, paced by the succession of measurement completions within their respective logoi. The substantial irreversibility of these events constitutes the authentic ``arrow of time'', which imprints information onto the object. The resulting physical phenomenon thereby realizes an \emph{ontological}, rather than merely epistemological, knowledge of the measured object.

Having established the general, discontinuous nature of measurement, it is instructive to examine the limiting case where this quantum tension dissolves. Thus, it is worthwhile to dwell on the ontology of measurement in the special case where the observational context is \emph{essential}, as introduced in Section~\ref{tevol}. In this case, the observables that define the universal context commute with the observational Hamiltonian defined in Eq.~\eqref{hobs}. If the state of the observed universe is prepared as an eigenstate of the observational Hamiltonian, this state remains stationary, undergoing no informational evolution and thereby preserving its existential tropos of actuality. The measurement of an observable within such a context neither alters the substantial element nor perturbs the tropos-existential condition of the measured system, and consequently, that of the observed universe as a whole. Ultimately, the essential context dissolves the ontological distinction between observation and measurement, as well as the tropos-existential distinction between environment and object. In this way, Bohrian epistemic objectivity is realized, simultaneously fulfilling the Einsteinian realism that ``allows the Moon to exist even when no one is looking at it''~\cite{Pais1982}.

\subsection{Criterion of objective reality}
\label{qucpr}

Before proceeding with a statement that reformulates the EPR criterion, adapting it to the quantum ontology outlined above, it is useful to briefly recapitulate the foundational ontological distinctions that emerged from the preceding discussion and, consequently, to make explicit the meaning of ``element of physical reality'', a notion left deliberately vague in the original EPR formulation, as discussed in Section~\ref{intro}. The emergent ontological constituents arising from the Hilbertian quantum formalism are classified as follows.
\begin{itemize}
\item \emph{Ontic}: Defined by mutually commuting classical states $\mathbb{P}_j$ corresponding to the substantial elements $\ket{\gamma_j}$. Operationally, they yield mutually exclusive outcomes associated with pre-existing definite values that are independent of observation. Their character is one of actuality and definiteness, proper to objective reality (both in the EPR and Copenhagen senses); the ontological regime corresponds to single-context or branch-wise definite states.
\item \emph{Processional}: Defined by mutually commuting observables corresponding to the substantial elements $\ket{\gamma_j}$. Operationally, they establish simultaneous measurability and enable the distinction between substantial elements and their respective physical manifestations, without altering the system's intrinsic character. Their character is actual, with a non-ontic distinction linking preparation to interrogation; the ontological regime is that of a fixed observational frame, wherein a change of substantial basis remains empirically vacuous within the given context.
\item \emph{Tropos-existential}: Defined by the non-commuting projectors $\mathbb{P}_j$ and $\mathbb{P}_k'$ within a contextual shift. Operationally, they give rise to interference and irreducible uncertainty, characterizing the unitary observational phase prior to the actualization of measurement. Their character combines actuality in preparation with potentiality and indeterminacy relative to the new frame; consequently, the ontological regime is multi-contextual and necessitates a non-unitary probabilistic update for objectification.
\end{itemize}
Thus, the statement of a ``criterion of physical reality'' provides a sufficient condition for the \emph{ontic} constituents of the Hilbertian quantum formalism -- i.e., the ``elements of physical reality''.

Building on these three ontological distinctions, we can now pinpoint where the original EPR criterion fails. Its conceptual limitations within the quantum domain stem from a category error: it attempts to apply a single-context, ontic framework to a multi-context, tropos-existential reality. Specifically, the EPR criterion:
\begin{enumerate}
\item fails to clarify the role of the overall observational context and does not explicitly articulate the environment-object distinction;
\item does not distinguish between the effects of the act of observation and those of measurement; in particular, it fails to differentiate the epistemic prediction of an environmental observable from the ontic determination of the object's physical manifestation;
\item does not account for potential characters inherent to non-objective reality, even when such characters lack ontic status.
\end{enumerate}
To overcome these conceptual limitations, we now reformulate the EPR criterion by explicitly integrating the ontological distinctions derived from the Hilbert-space framework. The revised statement preserves the conditional structure of the original EPR formulation but replaces its implicit non-contextual realism with a frame-relative ontological condition. By distinguishing between the observational environment and the measured object, and by recognizing the tropos-existential phase as a domain of potentiality rather than definiteness, the criterion treats descriptive objectivity as a sufficient condition for physical reality, thereby accommodating both actualized and non-objective aspects within a unified quantum ontology.

\begin{quote}
\textbf{Criterion of objective reality:} \emph{If, within a fixed universal context and without in any way disturbing the observational environment, we can predict with certainty (i.e., with probability equal to unity) the value of an environmental observable, thereby determining and ascertaining through measurement a physical manifestation of the object, then there exists an actual (ontic) character of the object corresponding to that observable}.
\end{quote}

While the preceding formulation establishes a general condition of objective reality valid across multi-contextual regimes, it admits a specialization in the case of an \emph{essential} universal context. Since in this regime the relevant observables commute with the observational Hamiltonian, the system's state remains stationary, and the ontological distinction between observation and measurement dissolves. Therefore, the criterion recovers the conditional structure of the original EPR statement, now grounded in a commutative frame rather than an implicit classical metaphysics.

\begin{quote}
\textbf{Criterion of essential reality:} \emph{If, without in any way disturbing a system, we can predict with certainty (i.e., with probability equal to unity) the value of an \emph{essential} observable, then there exists an actual (ontic) character of the system corresponding to that observable}.
\end{quote}

As noted in Section~\ref{intro}, Bohr remarked that the EPR criterion ``contains an essential ambiguity when it is applied to problems of quantum mechanics''~\cite{Bohr1935a}, precisely because it presupposes a context-independent notion of reality that ignores the constitutive role of the experimental arrangement. The present formulation addresses this ``essential ambiguity'' by explicitly restricting the domain of applicability to an essential context, where commutativity and stationarity guarantee that the condition of non-disturbance is not an idealization but a derivable property of the frame. In this way, the EPR intuition is recovered and validated, but it is confined to the domain where its logical premises are formally supported.

\section{Conclusion}
\label{concl}
Quantum theory does not stand in opposition to classicality; rather, it incorporates the classical description as its natural prerequisite. The ``classical world'' does not emerge merely as a dynamical physical regime wherein the quantum of action is negligible relative to the action of the system, but rather as an epistemic requirement that endows physical description with objectivity. Consequently, physical reality is inextricably bound to the classical structure as the sole framework capable of enabling the Boolean language necessary for its communicable description. Indeed, in regimes where the quantum of action is non-negligible, a non-Boolean, non-objective logic of physical reality emerges.

Since the von Neumann Hilbert-space formalism explicitly captures the relationship between quantum reality and the logic that describes it, it is natural in the present work to derive the physical condition for classicality by imposing the logical condition of Booleanity within the same formal framework. This yields a notion of classicality defined by the commutativity of preparable states, rather than that of observables: it is the states, and their mutual commutativity, that serve as the formal linchpin around which the entire quantum ontology revolves. Indeed, upon closer inspection, by demanding the simultaneous determination of properties on a remote system, the EPR authors effectively demanded the existence of a system state that is compatible with the observables corresponding to those properties, and thus the classicality of physical reality.

The logical requirement of classicality, imposed as the commutativity of physically preparable states, yields the HCM model. This control framework provides the formal description of the requisite environmental component in the process of observation. Just as a physical environment is indispensable to the act of observation, so too is Boolean logic, as a subalgebra of the quantum formalism, essential for the description of the observed object. In this manner, the observed reality -- the fundamental subject of a physical theory -- is invariably bipartite, divided into the environment and the object. The formal description of the former permits the inclusion of the observer within the theory of observation. Furthermore, the logical fluidity of their boundary affords the possibility of a consistent choice regarding the observed object. The experimental verification of physical decoherence constitutes the clearest expression of this freedom to define the demarcation.

Once the logical requirement of Booleanity is translated into the physical requirement of mutual compatibility among preparable states, thereby delineating the HCM formalism, it becomes possible to interpret this very formalism in ontological terms. This avoids the methodological error of imposing a ``classical metaphysics'' upon it a priori. Nevertheless, the resulting ontology exhibits pronounced affinities with the philosophical tradition:
\begin{itemize}
\item the deep, substantial level of physical reality is identified within the Hilbert space of states;
\item observables, represented as Hermitian operators on this space, formalize the existential (phenomenal) level;
\item classical states, described by fundamental propositional projectors on the same space, formalize the substantial-existential unity of the procession.
\end{itemize}
Further complementing these affinities is the tropos-existential connotation, formalized by the state as a general non-diagonal density matrix. This completes the quantum ontology and naturally yields a criterion of objective reality that is markedly contextual, reducing to the original EPR formulation only in the special case of an essential context.

In light of this, the present analysis reframes the historical EPR debate. It shifts the focus from the purported incompleteness of quantum mechanics to the hidden classical assumptions embedded within the EPR criterion itself. The methodological flaw in the EPR argument resides in the inversion of the relationship between formalism and ontology. Rather than allowing ontology to emerge a posteriori from the algebraic structure of a formalism that already successfully accounts for the phenomenology (whether objective or non-objective), EPR imposes a classical metaphysics a priori. The result is not a ``completion'' of quantum mechanics, but a category error that, as demonstrated by the reduction to the HCM model, impoverishes the theory by compelling it to relinquish the description of non-objective reality (interference), which is instead a constitutive and indispensable element of the quantum world. Thus, by showing that the joint requirements of locality, measurement independence, and the EPR criterion inevitably reduce the Hilbert-space formalism to the classical HCM model, we reveal that the EPR argument does not expose a deficiency in quantum theory, but rather precisely delineates the contextual boundaries of classical ontology. As a result, the foundational impasse can be addressed not by seeking ``hidden variables'', but by abandoning the traditional equivalence between epistemic objectivity and physical reality. The proposed criterion of objective reality supersedes the original EPR formulation by treating objectivity as simply a sufficient condition for reality, thereby accommodating the bipartite, contextual tropos of quantum phenomena.

Corresponding to the logico-physical bipartition between the environment and the object is a dual nature of information concerning the observed universe: it exhibits a fundamentally epistemic character, formally expressed through a Kolmogorovian probability with respect to the former, and an ontic character, formally expressed through a non-Kolmogorovian probability with respect to the latter. Epistemic probability quantifies the uncertainty pertaining to the environment, whereas ontic probability quantifies the indeterminacy pertaining to the observed object. Measurement is the culminating process that resolves this duality. By translating existential potentialities through the requisite classical interface, it converts them into objective, shareable knowledge for the observer via a conjoint act of epistemic revelation and ontic identification.

While the HCM model isolates the logical prerequisites of classicality, it abstracts away the dynamical richness of standard classical mechanics. Thus, this framework serves not as a final destination, but as a foundational stepping stone that may open several avenues for future research. Possible directions include:
\begin{itemize}
\item the explicit derivation of Hamiltonian or symplectic dynamics directly from the HCM framework, bridging the gap between its static logical skeleton and continuous temporal evolution;
\item the extension of the model to encompass realistic, non-ideal measurements, open quantum systems, and the foundations of quantum thermodynamics;
\item the treatment of dynamical chaos and the formulation of relativistic or quantum field-theoretic extensions of the classicality postulate;
\item the anchoring of such theoretical constructs to operational protocols and empirical tests of the quantum-classical boundary;
\item the application of this framework to refine decoherence theory, potentially via reformulations in Fock spaces.
\end{itemize}

Ultimately, pursuing these developments may solidify the unified, ontologically nuanced understanding of the quantum-classical transition advanced in the present work. Yet, the foundational takeaway remains epistemological. Indeed, the true mystery resides not in ``how the classical world physically emerges from the quantum realm'', but rather in recognizing a fundamental asymmetry: the classical cannot be derived from the quantum, because classicality -- understood as Boolean logic -- is the foundational logical prerequisite that allows the ``quantum'' to be articulated as a physical theory in the first place. Just as one cannot derive the rules of language from the physical phenomena it describes, one cannot derive the logical framework of classicality from the quantum formalism; rather, it must be presupposed as the very condition of possibility for any communicable and shareable physical description, and thus for the very possibility of knowledge itself.

\newpage

\appendix
\section{Ontological glossary}
\label{gloss}

\emph{Note}: The following glossary terms emerge \emph{a posteriori} from the Hilbert-space formalism constrained by the postulate of classicality. They recover a metaphysical vocabulary strictly as heuristic scaffolding to articulate the ontological structure that standard physical notation tends to flatten, without imposing external doctrinal commitments.

\begin{description}

\item[Character of the system:] The ontological unity binding a substantial element and an existential element, independently of the tropos-existential condition. Formally, it corresponds to the physical state $\rho$. \emph{Philosophical heritage}: Aristotelian-Thomistic substance-existence compound. (Sections~\ref{stobonto}, \ref{obsont})

\item[Classical basis:] The unique orthonormal eigenbasis $\{\ket{\gamma_i}\}$ diagonalizing all physically preparable states in HCM ($\rho=\sum_i p_i\ket{\gamma_i}\bra{\gamma_i}$). Defines the single classical observational context. (Section~\ref{stob})

\item[Classical information:] The Shannon entropy $I=-K\sum_i p_i\ln p_i$ quantifying epistemic ignorance in HCM. Vanishes upon measurement. \emph{Philosophical heritage}: Laplacian probability as incomplete knowledge. (Section~\ref{prob})

\item[Condition of observability:] A Hermitian operator $A$ qualifies as a physical observable only if it commutes with all preparable states ($[A,\rho]=0$). Ensures repeatability and context-definiteness. (Section~\ref{stob})

\item[Contextual shift:] A unitary transformation $U$ mapping one observational basis to a complementary, non-commuting one. Preserves the substantial element but alters the tropos-existential modality. (Sections~\ref{stob}, \ref{obsont})

\item[Element of physical reality:] In the original EPR formulation, it is a property predictable with certainty without disturbing the system. In the present quantum ontology, it is the \emph{ontic} constituents (actualized substantial elements) within a fixed or essential context, excluding tropos-existential potentialities. \emph{Philosophical heritage}: Aristotelian \emph{energeia} (actuality). (Sections~\ref{intro}, \ref{qucpr})

\item[Epistemic revelation:] The first phase of measurement; a Bayesian update of the environment's state via the L\"uders rule. Fixes the classical reference frame for objectification. (Sections~\ref{tevol}, \ref{meaonto})

\item[Essential context:] An observational context in which all defining observables are compatible with the observational Hamiltonian $\hat{H}_{\text{obs}}$. Yields stationarity, dissolving the conceptual distinction between observation and measurement. (Sections~\ref{tevol}, \ref{meaonto})

\item[Existential element:] An observable $\hat{O}$ manifesting a substantial element via the eigenstate-eigenvalue link. Ontologically (not ontically) distinct from the state. \emph{Philosophical heritage}: Aristotelian \emph{actus}. (Section~\ref{stobonto})

\item[Heisenberg cut:] The formal bipartition $\mathcal{H}_{\text{env}}\otimes\mathcal{H}_{\text{obj}}$ separating the classical environment from the quantum object. Epistemological, not macroscopic. \emph{Philosophical heritage}: Bohr's classical description requirement. (Section~\ref{obsont})

\item[Informational evolution:] The physical, non-epistemic variation of quantum information $I_{\text{obj}}(t)=-K\operatorname{Tr}(\rho_{\text{obj}}\ln\rho_{\text{obj}})$ during unitary observation. It reflects a tropos-existential evolution of the observed object, without altering the substantial element until measurement completion. (Sections~\ref{prob}, \ref{obsont})

\item[Logos of measurement:] A shared Boolean logical framework, instantiated by the observational environment, necessary for the unambiguous communicability of measurement outcomes and for circumventing self-referential paradoxes. \emph{Philosophical heritage}: Aristotelian \emph{logos} (rational account/structure). (Sections~\ref{meaonto}, \ref{qucpr})

\item[Measurement completion:] The joint event of epistemic revelation (environment) and ontic determination (object). Formalized by the L\"uders rule; induces discontinuous, irreversible state actualization and the arrow of time. (Section~\ref{meaonto})

\item[Observed object:] The quantum subsystem $\mathcal{H}_{\text{obj}}$ in tropos-existential potentiality. Evolves conditionally under $\hat{H}_{\text{obs}}$; non-objective until measurement. (Sections~\ref{stob}, \ref{obsont})

\item[Observational environment:] The classical subsystem $\mathcal{H}_{\text{env}}$ in existential actuality. Provides a fixed Boolean context for unambiguous communication. Modeled by HCM. (Section~\ref{obsont})

\item[Observational Hamiltonian:] The interaction operator $\hat{H}_{\text{obs}}=\sum_i\ket{\epsilon_i}\bra{\epsilon_i}\otimes\hat{V}^{(i)}$ ensuring conditional unitary evolution of the object per environmental branch. (Section~\ref{obsont})

\item[Observational universe:] A closed bipartite system $\mathcal{H}_{\text{env}}\otimes\mathcal{H}_{\text{obj}}$ evolving unitarily. Its dual tropos-existential structure prevents category errors leading to the measurement problem. (Sections~\ref{obsont}, \ref{meaonto})

\item[Ontic determination:] The second phase of measurement; the physical actualization of the character from potentiality to actuality. Replaces the substantial element probabilistically. (Section~\ref{meaonto})

\item[Ontic distinction:] The ontological distinction between classical states $\mathbb{P}_j\neq\mathbb{P}_k$. It corresponds to distinct substantial elements. \emph{Philosophical heritage}: Substance individuation. (Section~\ref{stobonto})

\item[Ontological distinction:] Umbrella term for ontic, processional, and tropos-existential distinctions, formalized via commutativity relations. It prevents category errors by assigning modal statuses to states and observables across different contexts. \emph{Philosophical heritage}: Aristotelian categorical distinctions. (Sections~\ref{stobonto}, \ref{obsont}, \ref{qucpr}) 

\item[Physical bootstrapping:] The mutual constraint between environment and object across the Heisenberg cut, experimentally probed via decoherence and cut mobility. (Section~\ref{obsont})

\item[Processional distinction:] The ontological, though not ontic, distinction between physical state and observable, or between observables sharing an eigenbasis. \emph{Philosophical heritage}: Thomistic real distinction between substance and existence. (Section~\ref{stobonto})

\item[Processional unity:] The ontological unity binding a substantial element and an existential element within a fixed context. Formally, it is expressed by the eigenstate-eigenvalue link $\hat{O}\ket{\gamma_j}\bra{\gamma_j}=o_j\ket{\gamma_j}\bra{\gamma_j}$. \emph{Philosophical heritage}: Aristotelian-Thomistic substance-existence compound. (Section~\ref{stobonto})

\item[Propositional projector:] An operator $\mathbb{E}_\Delta$ encoding experimental propositions $\mathfrak{p}_\Delta$. Mutual commutativity $\iff$ Boolean logic. Not identical to physical observables. (Section~\ref{class})

\item[Quantum information:] The von Neumann entropy $I=-K\operatorname{Tr}(\rho\ln\rho)$; a physical resource encoding contextual limits, preserved under unitary evolution. Non-epistemic. (Section~\ref{prob})

\item[Spatial non-contextuality:] The condition under which local marginals are independent of remote settings. Equivalent to local causality and the satisfaction of Bell inequalities; implies reduction to HCM. (Section~\ref{bell})

\item[Substantial element:] A basis vector $\ket{\gamma_j}$ anchoring ontic reality. Invariant under contextual shifts; bearer of properties. \emph{Philosophical heritage}: Aristotelian \emph{ousia}. (Section~\ref{stobonto})

\item[Tropos-existential actuality:] A mode of definite outcomes ($p=1$) within a fixed context. Characterizes the environment and essential eigenstates. \emph{Philosophical heritage}: Aristotelian \emph{energeia}. (Sections~\ref{tevol}, \ref{obsont})

\item[Tropos-existential distinction:] The ontological modal distinction between physical states of complementary contexts ($[\mathbb{P}_j,\mathbb{P}'_k]\neq0$), formalized by quantum probability. \emph{Philosophical heritage}: Aristotelian potency-act modal structure. (Sections~\ref{intro}, \ref{obsont}, \ref{meaonto})

\item[Tropos-existential potentiality:] A mode of indeterminate outcomes prior to measurement. Characterizes the observed object; non-objective until actualized. \emph{Philosophical heritage}: Aristotelian \emph{dynamis}. (Sections~\ref{obsont}, \ref{meaonto})

\item[Truth value:] A formal assignment $v_\rho(\mathfrak{p}_\Delta)=\operatorname{Tr}(\rho\mathbb{E}_\Delta)\in\{0,1,\text{indef.}\}$. Definite in HCM (Boolean); contextual in QM (orthomodular). (Section~\ref{class})

\end{description}


\begin{thebibliography}{99}

\bibitem{GreneNails1986}
Grene M. and Nails D. (eds.), \emph{Spinoza and the Sciences} (D. Reidel Publishing Company, Dordrecht, 1986).

\bibitem{Gabbey1996}
Gabbey A., Spinoza's natural science and methodology, in \emph{The Cambridge Companion to Spinoza}, edited by Garrett D. (Cambridge University Press, Cambridge, 1996).

\bibitem{Jammer1999}
Jammer M., \emph{Einstein and Religion: Physics and Theology} (Princeton University Press, Princeton, 1999).

\bibitem{BrewerWatkins2012}
Brewer K. and Watkins E., Difficulty still awaits: Kant, Spinoza, and the threat of theological determinism, \emph{Kant-Studien}, \textbf{103}(2): 163-187 (2012).

\bibitem{Melamed2017}
Melamed Y. Y., The causes of our belief in free will: Spinoza on necessary, ``innate'', yet false cognition, in \emph{Spinoza's Ethics: A Critical Guide}, edited by Melamed Y. Y. (Cambridge University Press, Cambridge, 2017).

\bibitem{EPR1935}
Einstein A., Podolsky B., Rosen N., Can quantum-mechanical description of physical reality be considered complete?, \emph{Phys. Rev.}, \textbf{47}: 777-780 (1935).

\bibitem{Bohr1949} 
Bohr N., Discussion with Einstein on epistemological problems in atomic physics, in \emph{Albert Einstein: Philosopher-Scientist}, edited by Schilpp P. A. (The Library of Living Philosophers, Evanston, 1949).

\bibitem{JahnertLehner2022} 
J\"ahnert M. and Lehner C., The early debates about the interpretation of quantum mechanics, in \emph{The Oxford Handbook of the History of Quantum Interpretations}, edited by Freire O. (Oxford University Press, Oxford, 2022).

\bibitem{Paty2022} 
Paty M., Einstein's criticism of quantum mechanics, in \emph{The Oxford Handbook of the History of Quantum Interpretations}, edited by Freire O. (Oxford University Press, Oxford, 2022).

\bibitem{Bohm1951}
Bohm D., \emph{Quantum Theory} (Prentice-Hall, New York, 1951).

\bibitem{BohmAharonov1957}
Bohm D. and Aharonov Y., Discussion of experimental proof for the paradox of Einstein, Rosen, and Podolsky, \emph{Phys. Rev.}, \textbf{108}: 1070 (1957).

\bibitem{Einstein1936}
Einstein A., Physik und Realit\"at, \emph{J. Frankl. Inst.}, \textbf{221}: 313-347 (1936).

\bibitem{Pais1982}
Pais A., \emph{Subtle is the Lord: The Science and the Life of Albert Einstein} (Oxford University Press, Oxford, 1982).

\bibitem{Bohr1935a}
Bohr N., Quantum mechanics and physical reality, \emph{Nature}, \textbf{136}: 65 (1935).

\bibitem{Bohr1935b}
Bohr N., Can quantum-mechanical description of physical reality be considered complete?, \emph{Phys. Rev.}, \textbf{48}: 696-702 (1935).

\bibitem{HarriganSpekkens2010}
Harrigan N. and Spekkens R. W., Einstein, incompleteness, and the epistemic view of quantum states, \emph{Found. Phys.}, \textbf{40}: 125-157 (2010).

\bibitem{PuseyBarrettRudolph2012} 
Pusey M. F., Barrett J., and Rudolph T., On the reality of the quantum state, \emph{Nat. Phys.}, \textbf{8}: 475-478 (2012).

\bibitem{Gleason1957}
Gleason A. M., Measures on the closed subspaces of a Hilbert space, \emph{J. Math. Mech.}, \textbf{6}: 885-893 (1957).

\bibitem{KochenSpecker1967}
Kochen S. and Specker E. P., The problem of hidden variables in quantum mechanics, \emph{J. Math. Mech.}, \textbf{17}: 59-87 (1967).

\bibitem{Barad2022}
Barad K., Agential realism -- A relational ontology interpretation of quantum physics, in \emph{The Oxford Handbook of the History of Quantum Interpretations}, edited by Freire O. (Oxford University Press, Oxford, 2022).

\bibitem{Folse1985}
Folse H. J., \emph{The Philosophy of Niels Bohr: The Framework of Complementarity} (North-Holland, Amsterdam, 1985).


\bibitem{vNeumann1955} 
von Neumann J., \emph{Mathematical Foundations of Quantum Mechanics}, translated by Beyer R. T. (Princeton University Press, Princeton, 1955).

\bibitem{Luders1950}
L\"uders G., \"Uber die Zustands\"anderung durch den Me{\ss}proze{\ss}, \emph{Ann. Phys. (Berlin)}, \textbf{8}: 322-328 (1950).

\bibitem{Koopman1931}
Koopman B. O., Hamiltonian systems and transformation in Hilbert space, \emph{Proc. Natl. Acad. Sci. U.S.A.}, \textbf{17}(5): 315-318 (1931).


\bibitem{KoopmanvNeumann1932} 
Koopman B. O. and von Neumann J., Dynamical systems of continuous spectra, \emph{Proc. Natl. Acad. Sci. U.S.A.}, \textbf{18}(3): 255-263 (1932).

\bibitem{Bondar2012}
Bondar D. I., Cabrera R., Loomis R. R., Rabitz H., Operator formulation of classical mechanics, \emph{Phys. Rev. A}, \textbf{86}: 022108 (2012).

\bibitem{Howard1994} 
Howard D., What makes a classical concept classical?, in \emph{Niels Bohr and Contemporary Philosophy}, edited by Faye J. and Folse H. J. (Kluwer, Dordrecht, 1994).

\bibitem{Bell1964}
Bell J. S., On the Einstein Podolsky Rosen paradox, \emph{Physics}, \textbf{1}(3): 195-200 (1964).

\bibitem{Kenny2006} 
Kenny A., \emph{A New History of Western Philosophy}, Vol. 3: \emph{The Rise of Modern Philosophy} (Clarendon Press, Oxford, 2006).

\bibitem{FernandezMoujan2024}
Fern\'andez Mouj\'an R., Greek philosophy for quantum physics. The return to the Greeks in the
works of Heisenberg, Pauli and Schr\"odinger, in \emph{Probing the Meaning of Quantum Mechanics}, edited by Aerts D., Arenhart J., de Ronde C., Sergioli G. (World Scientific, Singapore, 2024). 

\bibitem{Strumia2021}
Strumia A., A ``potency-act'' interpretation of quantum physics, \emph{J. Mod. Phys.}, \textbf{12}: 959-979 (2021).


\bibitem{PeresTerno2004}
Peres A. and Terno D. R., Quantum information and relativity theory, \emph{Rev. Mod. Phys.}, \textbf{76}: 93 (2004).

\bibitem{Strocchi2008}
Strocchi F., \emph{An Introduction to the Mathematical Structure of Quantum Mechanics}, Advanced Series in Mathematical Physics, Vol. 28 (World Scientific, Singapore, 2008).

\bibitem{BirkhoffvNeumann1936} 
Birkhoff G. and von Neumann J., The logic of quantum mechanics, \emph{Ann. Math.}, \textbf{37}(4): 823-842 (1936).

\bibitem{Jauch1968}
Jauch J. M., \emph{Foundations of Quantum Mechanics} (Addison-Wesley, Reading, MA, 1968).

\bibitem{FoulisBennett1994}
Foulis D. J. and Bennett M. K., Effect algebras and unsharp quantum logics, \emph{Found. Phys.}, \textbf{24}: 1325-1346 (1994).
 
\bibitem{BeltramettiCassinelli1981}
Beltrametti E. G. and Cassinelli G., \emph{The Logic of Quantum Mechanics} (Addison-Wesley, Reading, MA, 1981). 

\bibitem{Peres1984}
Peres A., What is a state vector?, \emph{Am. J. Phys.}, \textbf{52}: 644-650 (1984).

\bibitem{Peres2003}
Peres A., What's wrong with these observables?, \emph{Found. Phys.}, \textbf{33}: 1543-1547 (2003).

\bibitem{Fine1973} 
Fine A., Probability and the interpretation of quantum mechanics, \emph{Brit. J. Phil. Sci.}, \textbf{24}: 1-37 (1973).

\bibitem{Albert1992}
Albert D. Z., \emph{Quantum Mechanics and Experience} (Harvard University Press, Cambridge, 1992).



\bibitem{Kolmogorov1950}
Kolmogorov A. N., \emph{Foundations of the Theory of Probability}, translated by Morrison N. (Chelsea Publishing, New York, 1950).

\bibitem{Shannon1948} 
Shannon C. E., A mathematical theory of communication, \emph{Bell Syst. Tech. J.}, \textbf{27}(3): 379, 623 (1948).

\bibitem{Accardi1981}
Accardi L., Topics in quantum probability, \emph{Phys. Rep.}, \textbf{77}: 169-192 (1981).

\bibitem{BBCJPW1993}
Bennett C. H., Brassard G., Cr\'epeau C., Jozsa R., Peres A., Wootters W. K., Teleporting an unknown quantum state via dual classical and Einstein-Podolsky-Rosen channels, \emph{Phys. Rev. Lett.}, \textbf{70}: 1895-1899 (1993).


\bibitem{Thebault2021}
Th\'ebault K. P. Y., The problem of time, in \emph{The Routledge Companion to Philosophy of Physics}, edited by Knox E. and Wilson A. (Routledge, New York, 2021).

\bibitem{Rovelli1991}
Rovelli C., Time in quantum gravity: an hypothesis, \emph{Phys. Rev. D}, \textbf{43}(2): 442-456 (1991).

\bibitem{Barbour1994}
Barbour J. B., The timelessness of quantum gravity: I. The evidence from the classical theory, \emph{Class. Quantum Grav.}, \textbf{11}(12): 2853-2873 (1994).

\bibitem{Zurek2003} 
Zurek W. H., Decoherence, einselection, and the quantum origins of the classical, \emph{Rev. Mod. Phys.}, \textbf{75}: 715 (2003). 

\bibitem{Zurek2009}
Zurek W. H., Quantum Darwinism, \emph{Nat. Phys.}, \textbf{5}: 181-188 (2009).

\bibitem{Margenau1963} 
Margenau H., Measurements and quantum states, \emph{Philos. Sci.}, \textbf{30}(1): 1-16 (1963).

\bibitem{Earman1986}
Earman J., \emph{A Primer on Determinism} (D. Reidel Publishing Company, Dordrecht, 1986).


\bibitem{Saunders2018}
Saunders S., The Gibbs paradox, \emph{Entropy}, \textbf{20}(8): 552 (2018).


\bibitem{Howard1985}
Howard D., Einstein on locality and separability, \emph{Stud. Hist. Philos. Sci.}, \textbf{16}(3): 171-201 (1985).


\bibitem{Ballentine1970}
Ballentine L. E., The statistical interpretation of quantum mechanics, \emph{Rev. Mod. Phys.}, \textbf{42}: 358 (1970).

\bibitem{JWD2007}
Jones S. J., Wiseman H. M., Doherty A. C., Entanglement, Einstein-Podolsky-Rosen correlations, Bell nonlocality, and steering, \emph{Phys. Rev. A}, \textbf{76}: 052116 (2007). 

\bibitem{UCNG2020}
Uola R., Costa A. C. S., Nguyen H. C., G\"uhne O., Quantum steering, \emph{Rev. Mod. Phys.}, \textbf{92}: 015001 (2020).

\bibitem{Born1926} 
Born M., Zur Quantenmechanik der Sto{\ss}vorg\"ange, \emph{Z. Phys.}, \textbf{37}: 863-867 (1926).

\bibitem{Bell1966}
Bell J. S., On the problem of hidden variables in quantum mechanics, \emph{Rev. Mod. Phys.}, \textbf{38}: 447-452 (1966). 

\bibitem{Bohm1952}
Bohm D., A suggested interpretation of the quantum theory in terms of ``hidden'' variables, I and II, \emph{Phys. Rev.}, \textbf{85}: 166-193 (1952).

\bibitem{Bell1981}
Bell J. S., Bertlmann's socks and the nature of reality, \emph{J. Phys. Colloques}, \textbf{42}: C2-41 (1981).

\bibitem{Aspect1976}
Aspect A., Proposed experiment to test the nonseparability of quantum mechanics, \emph{Phys. Rev. D}, \textbf{14}: 1944 (1976). 

\bibitem{Lamehi-RachtiMittig1976}
Lamehi-Rachti M. and Mittig W., Quantum mechanics and hidden variables: a test of Bell's inequality by measurement of spin correlation in low-energy proton-proton scattering, \emph{Phys. Rev. D}, \textbf{14}: 2543 (1976).

\bibitem{Jarrett1984}
Jarrett J. P., On the physical significance of the locality conditions in the Bell arguments, \emph{No\^us}, \textbf{18}: 569-589 (1984).

\bibitem{Shimony1984}
Shimony A., Controllable and uncontrollable nonlocality, in \emph{Foundations of Quantum Mechanics in the Light of New Technology}, edited by Kamefuchi S. (Physical Society of Japan, Tokyo, 1984).

\bibitem{Fine1982}
Fine A., Hidden variables, joint probability, and the Bell inequalities, \emph{Phys. Rev. Lett.}, \textbf{48}: 291 (1982).

\bibitem{Bohr1928} 
Bohr N., The quantum postulate and the recent development of atomic theory, \emph{Nature}, \textbf{121}: 580-590 (1928).

\bibitem{Griffiths2002}
Griffiths R. B., \emph{Consistent Quantum Theory} (Cambridge University Press, Cambridge, 2002).

\bibitem{Lombardi2026}
Lombardi O., \emph{The Modal-Hamiltonian Interpretation of Quantum Mechanics: Making Sense of the Quantum World} (Oxford University Press, Oxford, 2026).

\bibitem{MKTKIWSMW2000}
Myatt C. J., King B. E., Turchette Q. A., Kielpinski D., Itano W. M., Wood C. J., Sackett C. A., Monroe C., Wineland D. J., Decoherence of quantum superpositions through coupling to engineered reservoirs, \emph{Nature}, \textbf{403}: 269-273 (2000).

\bibitem{HUHRBZA2003}
Hackerm\"uller L., Uttenthaler S., Hornberger K., Reiger E., Brezger B., Zeilinger A., Arndt M., Decoherence in a Talbot-Lau interferometer: the influence of molecular scattering, \emph{Appl. Phys. B}, \textbf{77}: 781-787 (2003).

\bibitem{FrauchigerRenner2018}
Frauchiger D. and Renner R., Quantum theory cannot consistently describe the use of itself, \emph{Nat. Commun.}, \textbf{9}: 3711 (2018).

\end{thebibliography}
\end{document}